\documentclass[aps,prd,english,superscriptaddress,11pt,notitlepage]{revtex4}
	\usepackage[colorlinks=true, a4paper=true, pdfstartview=FitV,
linkcolor=blue, citecolor=blue, urlcolor=blue]{hyperref}

\usepackage{amsmath}
\usepackage{amssymb}
\usepackage{graphicx}
\usepackage{hhline}
\usepackage{textcomp}
\makeatletter
\usepackage{babel}

\newcommand{\sech}{{\mathop{\mathrm{sech}}}}

\newcommand{\bea}{\begin{eqnarray}}
\newcommand{\eea}{\end{eqnarray}}

\newcommand{\be}{\begin{equation}}
\newcommand{\ee}{\end{equation}}
\usepackage[active]{srcltx}
\begin{document}


\title{Kink-antikink scattering in the $\phi^4$ model without static intersoliton forces }

\author{C. Adam}
\affiliation{Departamento de F\'isica de Part\'iculas, Universidad de Santiago de Compostela and Instituto Galego de F\'isica de Altas Enerxias (IGFAE) E-15782 Santiago de Compostela, Spain}
\author{K. Oles}
\affiliation{Institute of Physics,  Jagiellonian University, Lojasiewicza 11, Krak\'{o}w, Poland}
\author{T. Romanczukiewicz}
\affiliation{Institute of Physics,  Jagiellonian University, Lojasiewicza 11, Krak\'{o}w, Poland}
\author{A. Wereszczynski}
\affiliation{Institute of Physics,  Jagiellonian University, Lojasiewicza 11, Krak\'{o}w, Poland}

\begin{abstract}
 Kink-antikink scattering in the $\phi^4$ model is investigated in the limit when the static inter-soliton force vanishes. We observe the formation of spectral walls and, further, identify a new phenomenon, the vacuum wall, whose existence gives rise to a bouncing structure for the annihilating solitons. 
 
Furthermore, we discover higher order spectral walls, i.e., spectral walls which form when higher harmonics enter the continuous spectrum. These higher order spectral walls not only deform the soliton trajectories, they also can be observed easily as very intense radiation bursts. 
\end{abstract}

\maketitle

 \vspace*{0.2cm}
 
\section{Introduction}
The scattering of solitons and, especially, their annihilation, is a very complicated process which usually does not lead to integrable or solvable structures. There are several factors which contribute to the complexity of the
process. First of all, each soliton participating in the collision acts with a static force on the other participants, deforming their shapes \cite{MS}, \cite{shnir}. 
These deformations can also lead to the excitation of some internal modes, storing temporally some energy. The excited modes then can interact with the solitons and, as a consequence, influence the dynamics, e.g.,  via a so-called resonant mechanism. Finally, radiation can be produced in the scattering, which should be taken into account as it may have a nontrivial impact on the solitons, like, e.g., soliton creation \cite{tom-1} or negative radiation pressure \cite{tom-2}. All these types of interactions (by the static force, the internal modes and radiation) mix with each other, rendering any analytical treatment extremely difficult. As a consequence, there is still no satisfactory understanding of kink annihilation in various classical field theories, even on a qualitative level, see for example the problems in the $\phi^4$ model \cite{weigel}. 

Recently, a break-through in addressing this problem has been achieved, in terms of the so-called self-dual impurity (or background field) models \cite{BPS-imp}-\cite{BPS-imp-4}. These models provide the chance for a systematic study of  the interactions of solitons \cite{BPS-imp-solv}. They may be considered as a mathematical tool which allows, for a given theory and a given process, to completely switch off the static inter-soliton forces. Therefore, at leading order (small velocity of the asymptotically free kink-antikink pair), the annihilation process is described by a simple geodesic motion on a moduli space. However, these solitons do interact, mainly via the internal modes which can be excited in a controlled way. Hence, at next order, beyond the geodesic approximation, the annihilation is described by linear perturbation theory. This description is mathematically rigorous, because at each inter-soliton distance there exists a genuine static solution representing a kink-antikink pair. Thus, the modes are proper linear perturbation modes over a static BPS solution. This provides a very clear theoretical and numerical set-up for the study of soliton interactions. 

This linear perturbation theory beyond the geodesic approximation turned out to be sufficient to discover the spectral wall phenomenon \cite{spectral-wall}. A spectral wall is a spatially localized region, defined by the point(s) where an oscillation mode enters the continuous spectrum, which results in a nontrivial soliton interaction. 

Of course, in the final step the static force should be reintroduced, which implies a departure from the self-dual limit. Again, it can be performed in a perturbative, strictly controlled manner, which may enable us to disentangle which features are related to the internal modes and which appear rather as a result of the static forces. 

In the present work, we study the annihilation process in $\phi^4$ theory after coupling it with an impurity in such a way that the model remains self-dual in the topologically trivial channel, containing a kink-antikink pair. The aim is to understand this process beyond the geodesic flow approximation. In other words, we want to understand the impact of excitations of the internal modes on the soliton annihilation in a situation where the inter-soliton force is absent. Furthermore, we take some radiation effects into account, which also play a significant role. 

\section{Kink scattering in the $\phi^4$ model}

The investigation of the kink-antikink scattering in $\phi^4$ theory has a very long history, see \cite{moshir}-\cite{anninos}. $\phi^4$ theory is probably the most widely investigated soliton model, which serves as a prototypical non-integrable field theory with vast applications in many branches of physics \cite{kev-book}. It is a known fact that the soliton scattering in $\phi^4$ theory can result in many possible outcomes. For example, the initial solitons can be scattered back via an intermediate bound state in which the solitons suffer several bounces. The initial solitons can also annihilate into the vacuum, accompanied by the emission of radiation. This process can occur via the formation of an oscillon or of bions. All these possibilities give rise to a fractal-like structure. It is believed that the appearance of this structure is related to an internal bound mode (shape mode) which may store energy and transfer it to the translational degrees of freedom in a resonant way \cite{sugiyama}-\cite{weigel-2}. Although a complete quantitative or even qualitative description has not yet been provided \cite{kevrekidis-rev}, the role of the internal mode seems to be indisputable \cite{weigel}, \cite{weigel-3}. 

The problem with an analytical understanding of the kink-antikink scattering in the $\phi^4$ model is related to the non-existence of a moduli space representing this process, which means among other things, that there are no static kink-antikink solutions. By itself, this is not a particularly serious issue. It is known that in such a situation the moduli space can be replaced by an unstable manifold, which is an union of paths of steepest descent from a saddle point solution, which here is given by an infinitely separated pair of the kink $\phi_K$ and the antikink $\phi_A$ \cite{manton-unstabl}. Then, the simplest dynamics may be described by evolution on this unstable manifold via the collective coordinate method, i.e., by assuming that $\phi(x,t; x_0)=\phi_K(x, x_0(t))+\phi_A (x,-x_0(t))-1$, where $\pm x_0$ are the positions of the kink and antikink. However, the resulting effective model $$L[x_0]=\int dx \mathcal{L}_{\phi^4} [ \phi(x,t; x_0)]= M(x_0)\dot{x}_0^2(t) - V(x_0)$$ covers the elastic part of the scattering process only. Here $M(x_0)$ is a metric on the unstable manifold and $V(x_0)$ is an effective potential which leads to the appearance of a static intersoliton force. Note that the flow of the $x_0$ coordinate corresponds to the flow of the asymptotically zero mode of the free soliton, which now feels an attractive force.
 
In the next step, one should take into account the fact that the energy of the colliding solitons can be stored not only in the kinetic or potential part of the flow on the unstable manifold, but also in the massive mode, i.e., the shape mode, of the incoming solitons $A \eta(x,t)= A e^{-i\omega t} \sinh (x,x_0)/\cosh^2(x,x_0)$, where $\omega^2=3$ and $A$ is the mode amplitude. Hence, the approximated configuration is $ \phi(x,t; x_0, A_K,A_A)=\phi(x,t;x_0)+A_K\eta (x,t; x_0) +A_A\eta (x,t;-x_0)$, where the incoming solitons can be additionally boosted. The problem is that the obtained effective model $L[x_0, A_K, A_K; v]$ fails to reproduce the full dynamics.

One reason is that the unstable manifold we started with is a well defined approximation only if the energy of the initial $E_{in}$ and final $E_{out}$ states are comparable. But in the kink scattering $E_{in}=E_K+E_A$ while $E_{out}=0$, as the particles can annihilate to the vacuum. Thus, the unstable manifold is a valid approximation only for elastic processes. Even more importantly, the massive mode is affected by the presence of the second soliton. In fact, for $x_0=0$ we arrive at the $\phi=-1$ vacuum which does not support any modes at all. It means that the originally exited modes {\it disappear}. This fact is completely ignored by the collective coordinate approximation. 

Below we show that, in a limit where the moduli space does exist, the vanishing of the massive mode has a profound impact on the soliton dynamics in the kink-antikink scattering process. 
\section{Geodesic $K\bar{K}$ annihilation in the self-dual impurity $\phi^4$ model}
\subsection{The self-dual impurity models}

Here we briefly introduce the self-dual impurity models in (1+1) dimensions in the so-called solvable version \cite{BPS-imp-solv}. The theory is 
defined by the following total energy 
\be
 E=\int_{-\infty}^\infty  dx \left[\frac{1}{2}\phi_t^2+ \frac{1}{2} \phi_x^2  + W^2 + \sigma^2 W^2 +\sqrt{2} \sigma W \phi_x + 2 W^2 \sigma \right]
 \ee
 where $W$ is a prepotential (a function of the real field $\phi$) and $\sigma$ is an impurity (a background field).  In the limit when the impurity identically vanishes, $\sigma \equiv 0$, the model tends to the standard scalar field theory with a potential $U=W^2$. Of course, in oder to deal with topological solitons one should assume that the potential possesses at least two vacua at $\phi_+ > \phi_-$. Then, the kink (antikink) is a solution with a positive (negative) value of the topological charge $Q$ interpolating between $\phi_-$ and $\phi_+$ ( $\phi_+$ and $\phi_-$). The important fact is that this coupling of the impurity leads to a self-dual (BPS) static sector. Indeed, the static energy can be rewritten as 
  \bea
 E=\int_{-\infty}^\infty  dx \left[ \frac{1}{2} \left( \phi_x +\sqrt{2} W \right. \right. &+& \left. \sqrt{2} \sigma W \right)^2  - \left. \sqrt{2} W \phi_x \right] \nonumber \\
 &\geq&  - \sqrt{2} \int_{-\infty}^\infty dx W \phi_x = -\sqrt{2} Q \int_{\phi_-}^{\phi_+}W d\phi
 \eea
where the bound is saturated if the following Bogomolny equation is satisfied 
\be
 \phi_x +\sqrt{2} W + \sqrt{2} \sigma W=0. \label{BOG-x}
\ee
One can check that the Bogomolny equation implies the static Euler-Lagrange equation. The existence of the Bogomolny equation implies the existence of infinitely many static solutions which form a moduli space $\mathcal{M}$, i.e., a space of energetically equivalent solutions parametrized by a coordinate on $\mathcal{M}$. In the no-impurity limit, this space is just a space of a kink located at an arbitrary position. Thus, the transition from one BPS solution to another is given by a translation. This results in a trivial metric on $\mathcal{M}$. Also the spectral structure is independent of the position on the moduli space. In the self-dual impurity model, the form of the solution as well as its spectral structure {\it do depend} on the position on the moduli space. This is a crucial feature which makes the self-dual soliton impurity models very useful tools for studying interactions of topological solitons. 

\subsection{The topologically trivial BPS sector and geodesic motion}

\begin{figure}
\hspace*{-1.0cm}
\includegraphics[height=4.5cm]{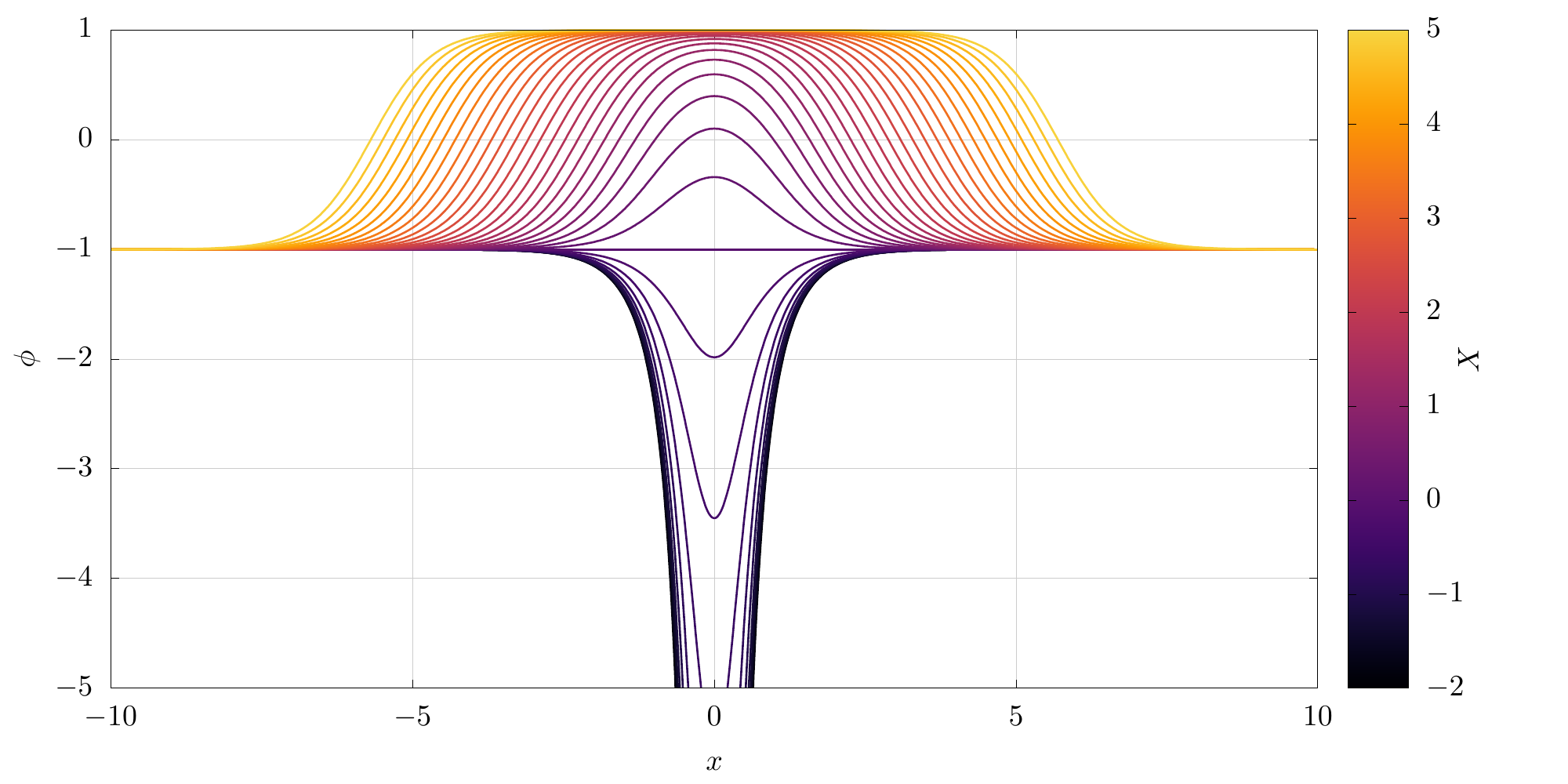}
\includegraphics[height=4.5cm]{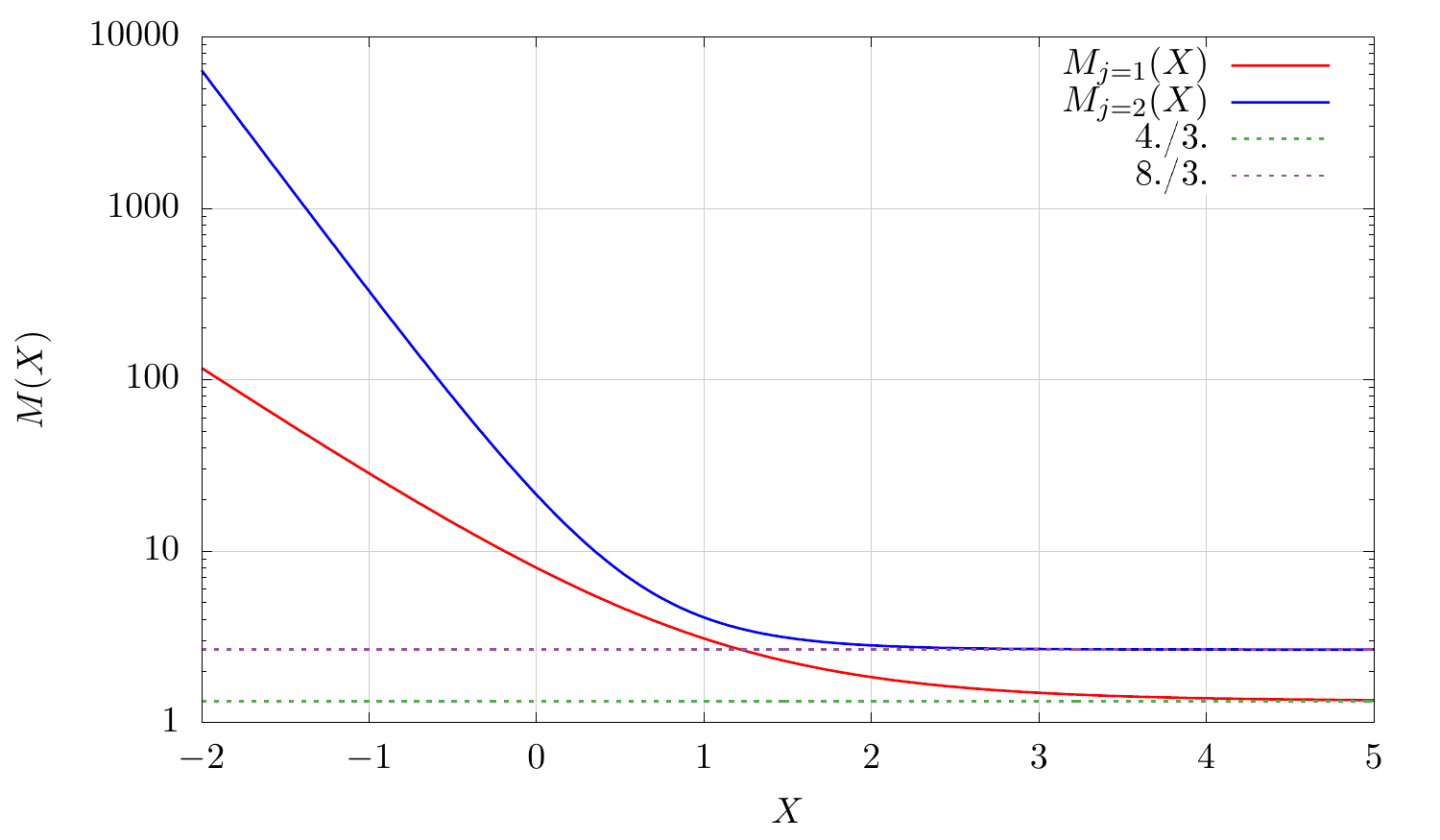}
\caption{{\it Left}: BPS solutions for different values of the moduli parameter $X$ for $\sigma_{j=1}$. {\it Right}: Metric on the moduli space. The horizontal lines denote the asymptotic values $M(X=\infty)$.}
\label{BPS-plot}
\end{figure}

In the present work we choose the prepotential as
\be
W=\frac{1}{\sqrt{2}} (1-\phi^2) .
\ee
Then the model in the no-impurity limit reduces to $\phi^4$ theory. Furthermore, in order to study the annihilation process, we use the following non-localized impurity
 \be
\sigma_j= \frac{j}{2} \tanh x-1
 \ee
 where $j \in \mathbb{R}$ is a parameter which measures the strength of the impurity and controls the mode structure. The BPS  sector consists of the topologically trivial solutions 
\be
\phi(x; a)= - \frac{\cosh^j x -a}{\cosh^j x +a}
\ee
where $a\in (-1, \infty)$ is a moduli space coordinate. It is also useful to introduce another moduli space coordinate $X \in \mathbb{R}$, such that
\be
a=-1+e^{jX}.
\ee
These BPS solutions obey the pertinent Bogomolny equation and have zero energy, because they saturate a topological energy bound. Observe that the solutions with $X\geq 0$ correspond to $\tanh$ antikink (kink) solutions, while for $X <0$ to $\coth$ singular solutions of the original (no impurity) theory \cite{BPS-imp-solv}. In Fig. \ref{BPS-plot}, left panel,  we present the BPS solutions for different values of the moduli coordinate $X$ for the $\sigma_{j=1}$ impurity. It is clearly visible that for $X\rightarrow \infty$ it represents an infinitely separated kink-antikink pair of $\phi^4$ (with mass threshold $j^2$). As we tend to $X=0$, the solitons approach each other and annihilate, forming  the $\phi=-1$ lump (which is a counterpart of the vacuum solution in the no-impurity model). For negative $X$, the BPS solution develops a well whose bottom tends to an arbitrary negative value as $X \rightarrow - \infty$. 

In the limit of small kinetic energy, it is plausible to expect that the true evolution follows a sequence of energetically equivalent BPS solutions. To find such a flow on the space of BPS solutions, i.e., on the moduli space, one promotes the moduli coordinate to a time dependent variable. Then, after spatial integration, we derive an effective theory
\be
L=-\frac{1}{2} M(a) \dot{a}^2
\ee
where the metric (mass) on the moduli space reads
\be
M_j(a)=\int_{-\infty}^\infty dx \left( \frac{d}{da}  \phi(x; a) \right)^2 = 4 \int_{-\infty}^\infty \frac{\cosh^{2j} x }{(\cosh^j x +a )^4} dx .
\ee
For $j=1$, the metric can be found in an analytical form 
\be
M_{j=1}(a)= -\frac{4}{3} \frac{1}{(1-a^2)^3} \left( a^4-10a^2-6 - 6a\frac{4+a^2}{\sqrt{1-a^2}} \arctan \frac{a-1}{\sqrt{1-a^2}} \right) .
\ee
It is a general feature that the metric is a well-behaved function for any value of the moduli space coordinates. Only at the left boundary ($X\rightarrow -\infty$ i.e., $a \rightarrow -1$) it approaches infinity. However, this happens in infinite time \cite{BPS-imp-solv}. Note also that at $X=0$ ($a=0$), i.e., where the solution goes through the lump form $\phi=-1$, the metric has a finite value. Hence, from the point of view of the geodesic motion, this is not a special point. In our numerical simulations, we compare the geodesic approximation with the true dynamics given by the full equation of motion
\be
\phi_{tt}-\phi_{xx} - \frac{j}{2}(1-\phi^2) \frac{1}{\cosh^2 x} \left( 1+j \phi  \sinh^2 x\right)=0
\ee
for an initially widely separated kink-antikink pair with small velocity. As expected, we observe a very good agreement between these two descriptions. 
\section{Beyond geodesic approximation: spectral walls}
\subsection{Flow of the spectral structure on the moduli space}

\begin{figure}
\hspace*{-1.0cm}
  \includegraphics[height=5.0cm]{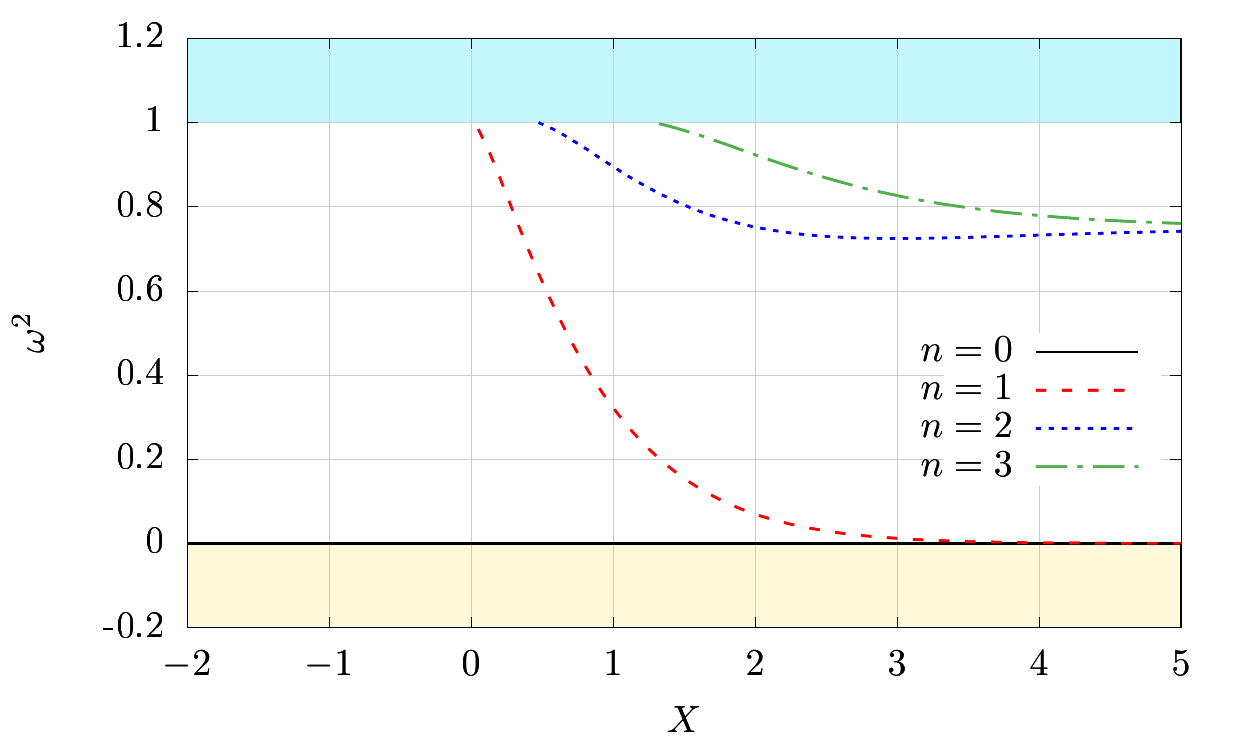}
  \includegraphics[height=5.0cm]{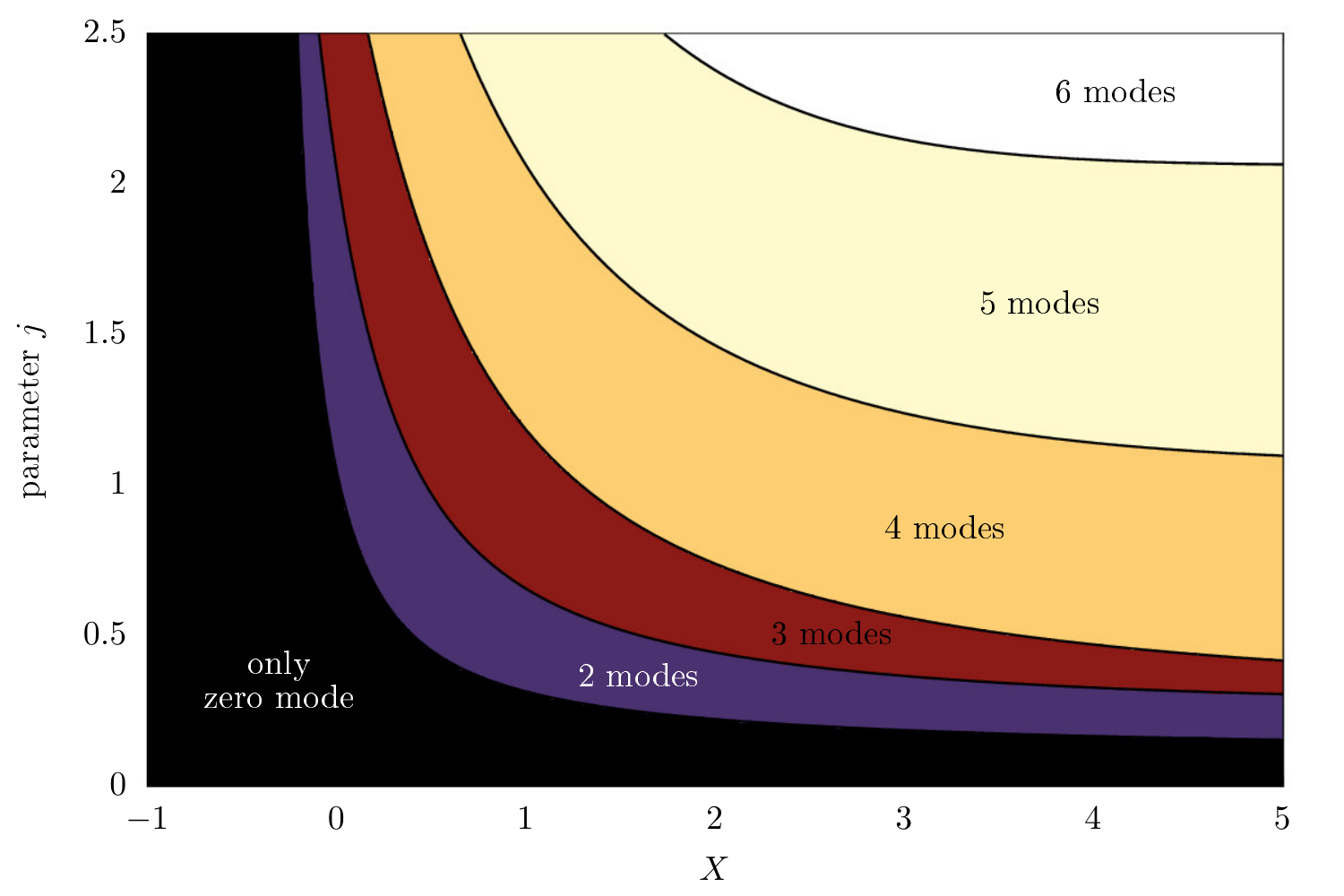}
\caption{{\it Left}: Dependence of the spectral structure on the position on the moduli space $X$ for $\sigma_{j=1}$. {\it Right}: Dependence of the spectral structure on the position on the moduli space $X$ for different $j$.}
\label{modes-plot}
\end{figure}

The first correction to the geodesic flow on the moduli space is given by the linear perturbation theory, that is by an inclusion of the lightest modes (i.e., the bound modes), on top of the zero mode.  

The spectral structure of the BPS solutions asymptotically (for $X\rightarrow \infty$) decomposes into a sum of two $\phi^4$ theories with the mass threshold at $E=j^2$ (two zero modes plus two discrete modes) and an additional number of discrete modes introduced by the lump solution $\phi=+1$. Of course, as is typical for the self-dual soliton impurity models, the mode structure changes while we move on the moduli space - see Fig. \ref{modes-plot}, let panel, where the spectral structure for the $\sigma_{j=1}$ impurity as a function of the moduli space coordinate is presented. In Fig. \ref{modes-plot}, right panel, we show the number of modes for $j \in (0,2.5)$ as a function of $X$. 

Let us discuss the $j=1$ case in detail. At $X\rightarrow \infty$ we have two bound modes which are exactly the bound (shape) modes of the $\phi^4$ kink and antikink. There are also two zero modes, as for an infinitely separated pair of solitons each of them can be moved independently without any cost of energy. As $X$ takes a finite positive value, the constituent solitons interact and only one zero mode remains. This mode, interpreted as a symmetric change of the position of solitons, provides the moduli space flow. The antisymmetric counterpart becomes a massive bound mode which hits the continuous spectrum (as the last mode) at $X=0$. Two $\phi^4$ bound modes also mix, forming a symmetric (antisymmetric) superposition with lower (higher) energy. The antisymmetric mode enters the continuous spectrum at $X=1.25$ and the symmetric mode at $X=0.39$.  It is convenient to relate the moduli space coordinate to the value of the field at the origin 
\be
\phi_0 \equiv \phi(x=0) = \frac{a-1}{a+1}=1-2e^{-jX}  \;\;\; \Rightarrow \;\;\; X=\frac{1}{j} \ln \frac{2}{1-\phi_0}
\ee
To summarize, we find three bound modes which enter the continuous spectrum. Therefore, we expect three spectral walls for: 1) the antisymmetric shape mode superposition at $\phi_0\approx 0.36$, 2) the symmetric shape mode superposition at $\phi_0 \approx -0.34$ and 3) the antisymmetric translational mode superposition at $\phi_0 =-1$. 
\begin{figure}
\hspace*{-1.0cm}
  \includegraphics[height=5.5cm]{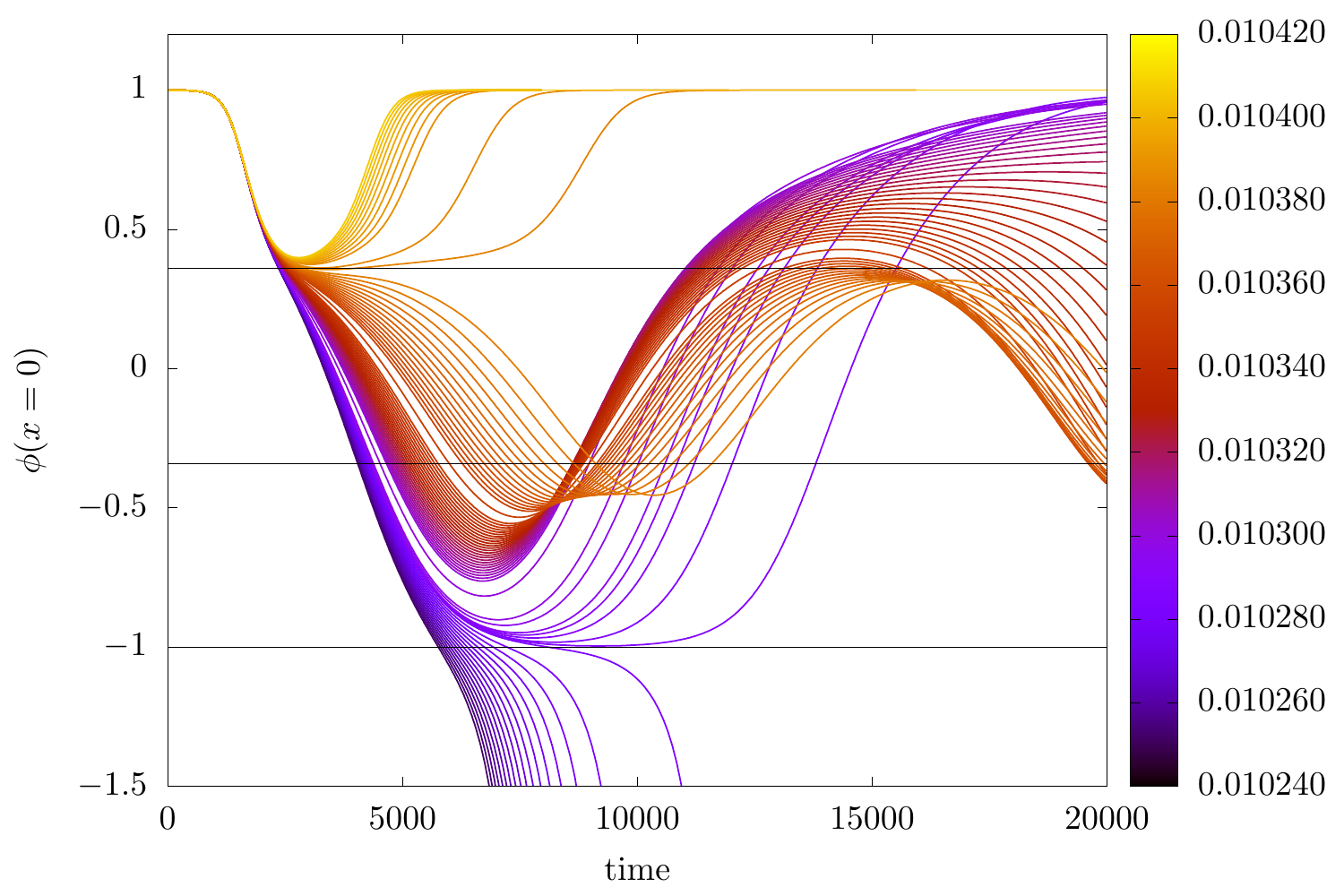}
      \includegraphics[height=5.5cm]{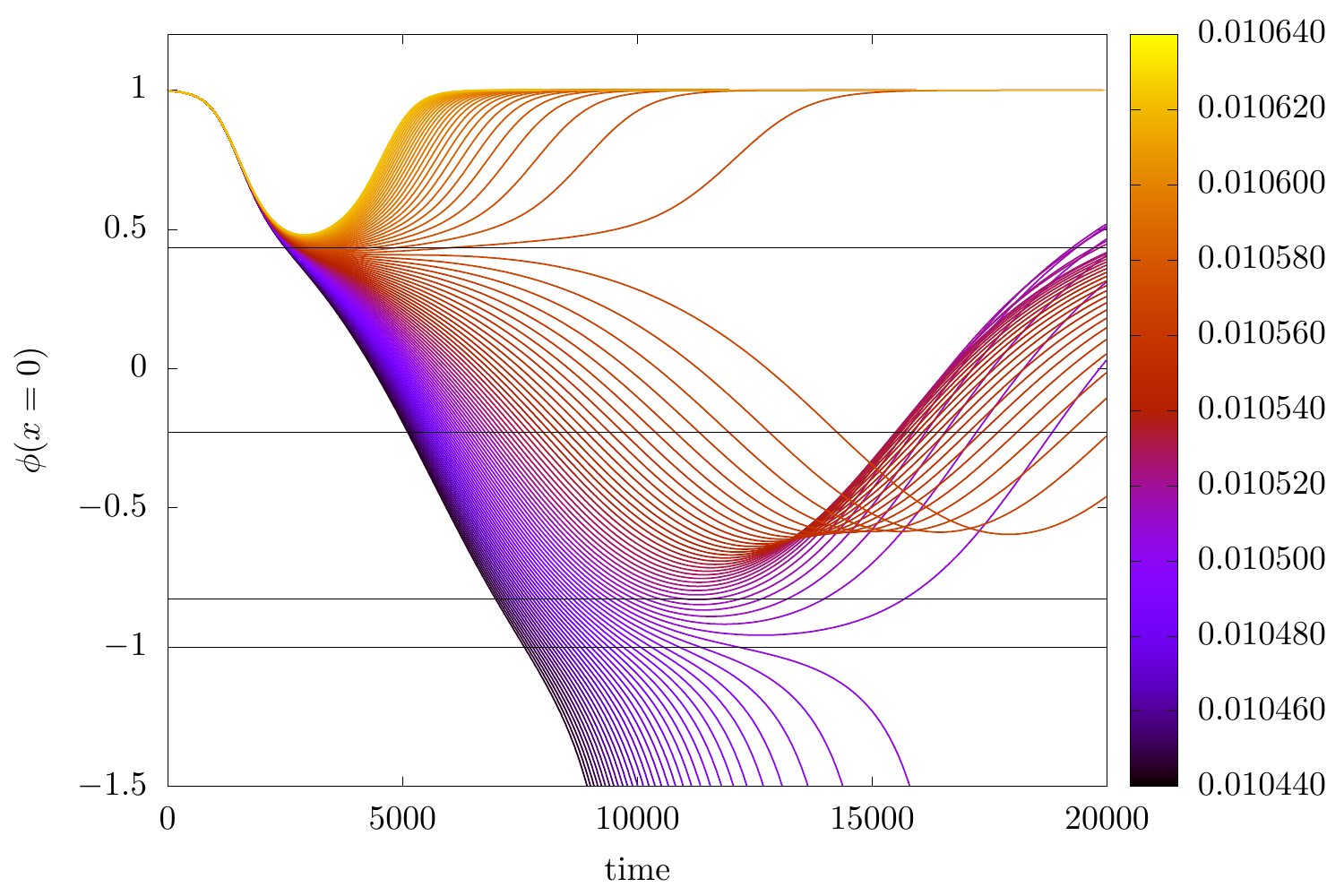}
\caption{The value of the field at the origin $\phi_0\equiv \phi(x=0)$ for the asymmetric superposition of the shape modes excited. The incoming  solitons speed is $v=0.005$. {\it Left}: Impurity $\sigma_{j=1}$.  {\it Right}: Impurity $\sigma_{j=0.7}$.}
\label{asym-wall-plot}
\end{figure}

\subsection{Spectral walls}
To verify the existence and analyze the properties of the spectral walls, we perform numerical simulations of the full equation of motion. As there are three bound modes the system may allow for interactions between the spectral walls.  

We start with the spectral wall located at $\phi_0=0.363910$. It is triggered by the antisymmetric superposition of the shape modes of the incoming, large separated kink and antikink which are then boosted towards each other
\bea
\phi_{in}(x,t) =&-& \tanh\left(\frac{j\gamma}{2} (x-x_0-vt) \right) - A_1 \frac{\sinh \left(\frac{j\gamma}{2} (x-x_0-vt) \right) }{\cosh^2 \left(\frac{j\gamma}{2} (x-x_0-vt) \right) } \cos \omega_j\gamma(t-vx) \nonumber \\
&+& \tanh\left(\frac{j\gamma}{2} (x+x_0+vt) \right) + A_2 \frac{\sinh \left(\frac{j\gamma}{2} (x+x_0+vt) \right) }{\cosh^2 \left(\frac{j\gamma}{2} (x+x_0+vt) \right) } \cos \omega_j\gamma(t+vx)-1 \label{init-pert}
\eea 
where $A_1=-A_2\equiv A$, $\omega_j^2=3j^2/4$ and $x_0 \gg 1$. 

This case is presented in Fig. \ref{asym-wall-plot}, left panel. The corresponding wall is very well visible. As expected for a sufficiently large excitation of the mode, the incoming kink-antikink configuration is reflected back. For a smaller (critical) amplitude, the incoming configuration gets stuck on the wall precisely at the value of the moduli space coordinate $\phi_0$ where the relevant mode enters the continuous spectrum. For even smaller excitations, the configuration more and more freely passes the wall. However, $\phi(x=0)$ does not tend to an arbitrary value as one would expect from the geodesic flow. In fact, there are two new phenomena which strongly modify the dynamics of the kink-antikink pair. 

First of all, the incoming configuration (after passing through the wall) feels another wall at $\phi_0=-1$. The observed reflection at the point $\phi_0=-1$ would not be too surprising. It corresponds to a spectral wall related to a different asymmetric mode, that is, the asymmetric superposition of the translational modes of the incoming solitons.  It could be perfectly possible that energy from one asymmetric mode is transferred to another asymmetric mode. Indeed, such an energy transfer exists and is related to another property of the spectral wall. 
To better understand the problem, we consider the same set-up for the $j=0.7$ impurity, for which no spectral wall coincides with $\phi_0=-1$. However, once again we see a reflection from $\phi_0=-1$, see Fig. \ref{asym-wall-plot}, right panel. Apparently the point $\phi_0=-1$, although being an ordinary point of the moduli space metric, acts as a sort of a new type of wall, which is not related to any specific internal mode. On the contrary, as we will see below, it reflects any state with a sufficiently large mode. This should be contrasted with the spectral walls which, being rather selective objects, are transparent if the corresponding mode is not excited. 

Note that the wall for $\phi_0=-1$, let us call it the {\it vacuum wall} as the field configuration corresponds to one of the vacua of the original $\phi^4$ model, is responsible for the fact that a back scattered trajectory may give rise to {\it bouncing solutions}. This is precisely what we observed for some solutions - see orange trajectories in Fig. \ref{asym-wall-plot}, left panel. However, the reflected back solution {\it is not} repelled again by the original spectral wall at $\phi_0\approx 0.36$. As we explain below, there is another mechanism, related to the existence of the so-called {\it higher spectral walls}, which is responsible for this behavior. Here we conclude that we do have a bouncing structure in the kink-antikink annihilation only via the mode interaction triggered by three necessary ingredients: the spectral wall (at which energy from the originally excited mode is transferred to other modes), the vacuum wall and higher order spectral walls. It should be underlined that in the case at hand this is a mathematically rigorous explanation. That is to say, the static solutions (moduli space solutions), above which the bound modes are found, exist for any distance between the kink and the antikink. 
  \begin{figure}
  \includegraphics[height=6.0cm]{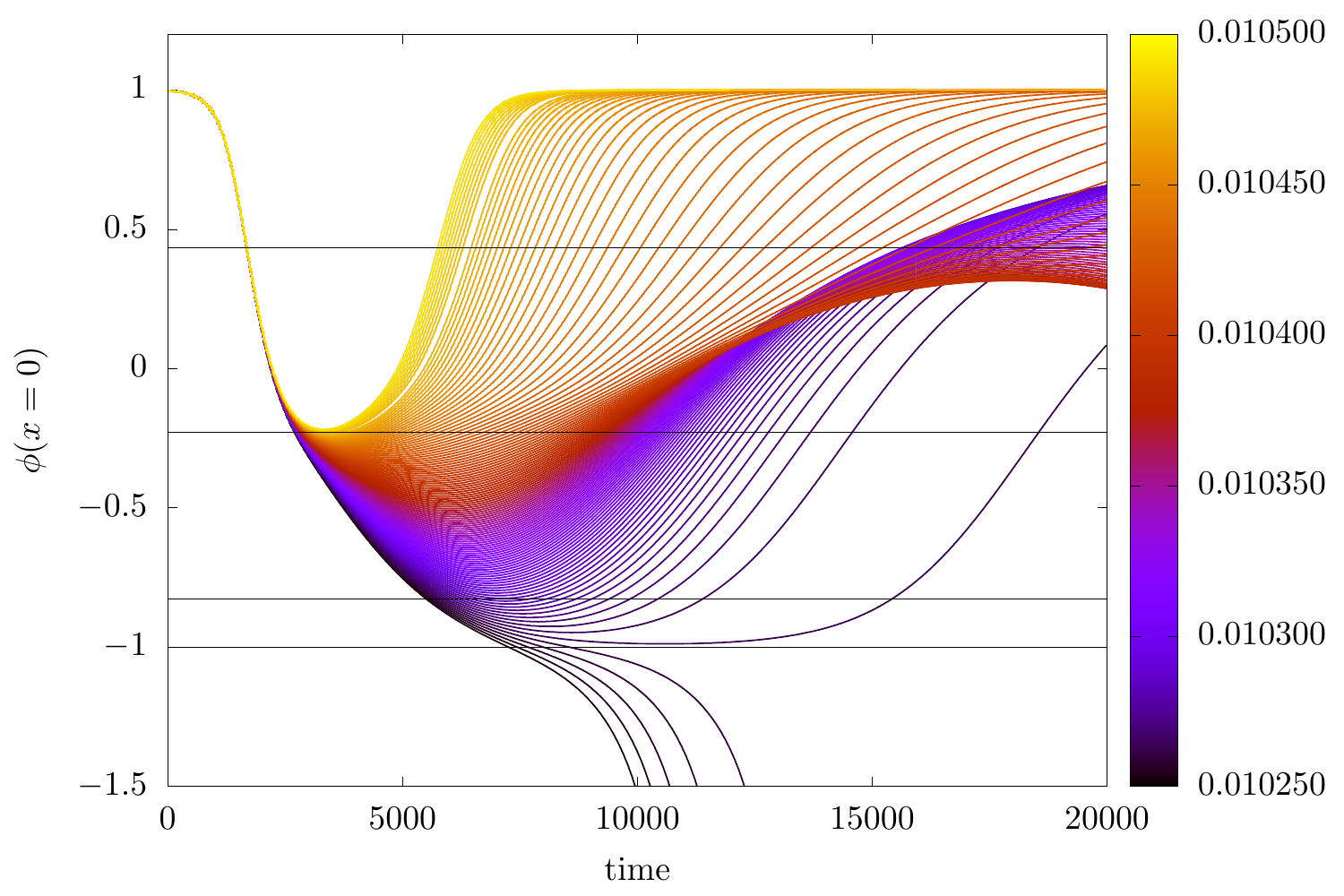}
\caption{The value of the field at the origin for the symmetric mode excited. The incoming  solitons speed is $v=0.005$ and impurity $\sigma_{j=0.7}.$ The spectral walls are located at $ \phi_0= 0.432919, -0.228658,  -0.827630$.}
\label{sym-wall-plot}
\end{figure}

Finally, for smaller amplitudes even the vacuum wall is traversable as dictated by the geodesic approximation. 

\vspace*{0.2cm}

The whole analysis can be repeated when the symmetric superposition of the shape modes is excited, $A_1=A_2=A$. Then, for $j=0.7$ one expects a spectral wall at $\phi_0=-0.228658$. This situation is plotted in Fig. \ref{sym-wall-plot}. It turns out that this spectral wall is not as well visible as the previous one. There is a range of amplitudes for which the trajectory is attracted by the $\phi_0=-0.23$ region, see the orange curves in Fig. \ref{sym-wall-plot}. However, we did not find a trajectory (value of the amplitude of the initial perturbation) which would be attached to this spectral wall for a longer time. A reason for that could be the close presence of the vacuum wall at $\phi=-1$. Hence, each solution which would go through this spectral wall is immediately repelled by the vacuum wall. It could be that even at the spectral wall this repulsion does play a role. However, this issue requires further studies. 

Note that also in this case there are solutions which, after reflection from the vacuum wall, are once again reflected which results in a bouncing solution, similarly to those found where the asymmetric mode was excited. 

\vspace*{0.2cm}

Undoubtedly, the vacuum wall has a different origin than the spectral walls. Its source is not any bound mode entering the continuous spectrum. As a consequence, also its behavior is quite distinct as it affects {\it any} excited BPS solution. As we saw, this may have a non-trivial impact on nearby spectral walls. 
 
\subsection{The effective model}
\begin{figure}
 \includegraphics[width=\textwidth]{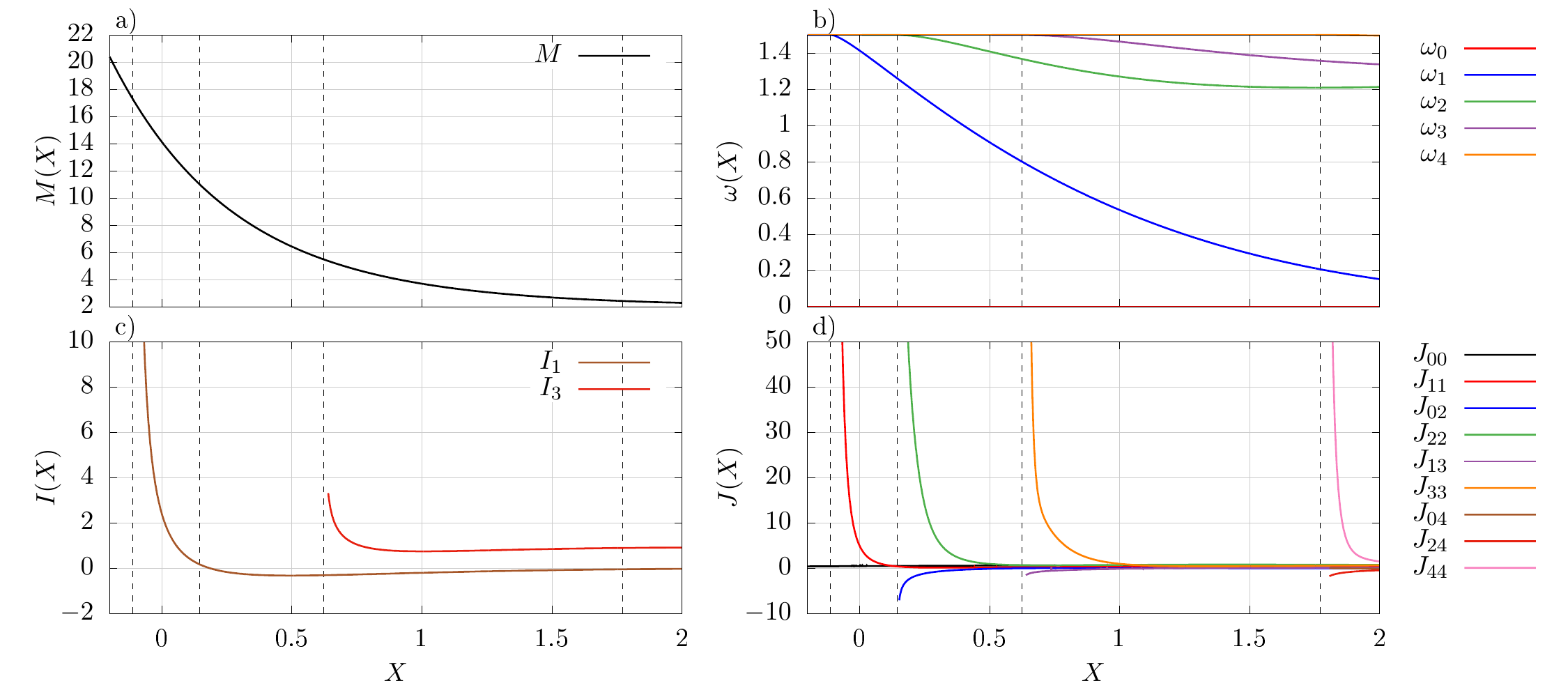}
 \caption{Effective model for $j=1.5$: {\it a)} effective mass;  {\it b)} frequencies; {\it c)} non identically vanishing integrals I and {\it d)} $J$. }
 \label{eff}
\end{figure}
The existence of the spectral walls can be further explained in terms of an effective model. To construct such an  effective model, we take into account the zero mode, providing the lowest order geodesic flow, as well as the lightest bound modes 
\begin{equation}
 \phi=\Phi(x, X(t))+A_i(t)\eta_i(x,X(t))
\end{equation} 
where the moduli space coordinate $X$ and the amplitudes $A_i$ of the modes are promoted to time dependent variables. We underline that, in contrast to the BPS solution $\Phi$ which is known exactly, the discrete modes have to be computed numerically at any value of the moduli coordinate $X$. This is a very nontrivial task, which requires to solve the perturbation problem (Schrodinger equation) at any value of the moduli space coordinate. Finally, after integrating over the spatial dimension, we arrive at the following effective Lagrangian
\begin{equation}
 L=\frac{1}{2}M\dot X^2+\frac{1}{2}\dot A_i\dot A_i+I_iA_i\dot X+J_{ik}A_iA_k\dot X^2-\frac{1}{2}\omega_i^2A_i^2+\mathcal{O}(A^3)
\end{equation} 
where in addition to the metric on the moduli space 
\begin{equation}
 M(X) = \int \Phi_X^2
\end{equation} 
we define the following mode coupling functions
\begin{equation}
  I_i(X) = \int \eta_{i,X}\Phi_X,\qquad J_{ik}(X)=\int \eta_{i,X}\eta_{k,X}
\end{equation} 
which describe the mutual interaction between the zero mode and the massive bound modes.
Note that due to the symmetry $I_i\equiv 0$ for even $i$. Moreover, $J_{ik}\equiv 0$ if $i+k$ is an odd number.
As expected, the integrals $I_i$  and $J_{ik}$ are singular at the appropriate walls. Specifically, $I_i$ ($i$ odd) and $J_{ii}$ tend to $+\infty$ at the corresponding walls, whereas  $J_{ik}$ to $-\infty$ if $i\neq k$ - see Fig. \ref{eff}. Therefore, the positions of the spectral walls can be deduced from the points of break-down of the effective model, i.e., where the pertinent integrals diverge.

\section{The vacuum wall}
To explain the existence of the vacuum wall, or, equivalently, the special character of the $\phi=-1$ BPS solution, we have to go to the next order of perturbation theory. However, a heuristic argument may be formulated as follow. When we added the initial perturbations, we did it on a BPS solution with a large value of the variable $X$ on the moduli space. Thus, the perturbation is added on each of the constituent solitons separately. Each soliton is a massive (heavy) particle, with a nontrivial energy density distribution, and the perturbation weakly changes its mass. Then,  we evolve the initial configuration, i.e., send the perturbed kink and antikink towards each other. Once we approach $X=0$, the individual solitons disappear and we arrive at the uniform  $\phi=-1$ solution.  This configuration belongs to the moduli space and so has zero static energy, but now it also has zero static energy density (by "static" we mean that the kinetic contribution $\sim \dot \phi^2 $ is not taken into account). But this implies that now we reached a point at which the system is very sensitive to any perturbation. From the linear analysis we know that there exists a zero mode governing the flow on the moduli space. This mode evolves and becomes either two kinks going in opposite directions $(X>0)$ or collapses into the singularity at infinite time $(X<0)$. However, the model is nonlinear and nonlinearities, accessible at higher order perturbation theory, can give a contribution and push the zero mode in one of the possible directions. 

\subsection{Numerical simulations}
\begin{figure}
 \includegraphics[height=4.8cm]{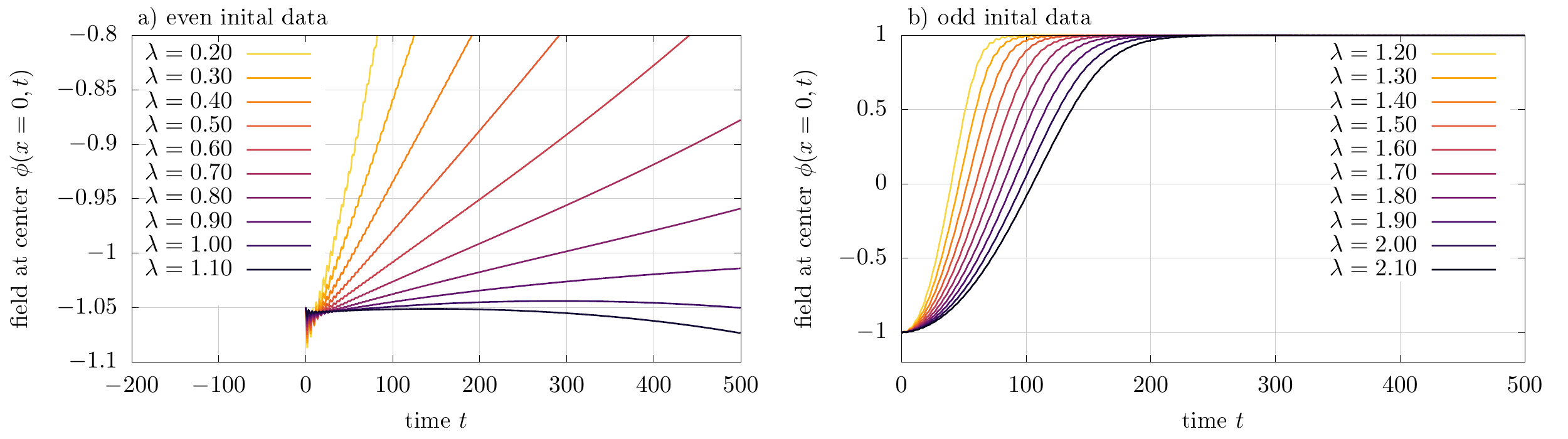}
 \caption{Evolution of the system for even and odd initial data for various lambdas and $A=-0.05$ and $j=1.5$.}\label{figPerturbations}
\end{figure}

We studied both even and odd initial conditions imposed on top of the $\phi=-1$ BPS solution. For the even perturbation we assumed 
\begin{equation}
 \phi(x,t=0)=-1+\frac{A}{\cosh^\lambda x},\qquad \phi_t(x,t=0)=0,\qquad \lambda>0.
\end{equation} 
For $\lambda=j$ such data coincide with the profile of the zero mode.
For the odd perturbation 
\begin{equation}
 \phi(x,t=0)=-1+A\frac{\sinh x}{\cosh^\lambda x},\qquad \phi_t(x,t=0)=0,\qquad \lambda>1.
\end{equation} 
The results are gathered in Figure \ref{figPerturbations}. For $A>0$, in both cases the system evolves into a pair of kinks, i.e., we move toward positive $X$ on the moduli space. For $A<0$, in the examined range the odd initial data also evolved always into the creation of the $K\bar K$ pair. Even initial data for $A<0$ bifurcate. There is some critical $\lambda$ (in our case $\lambda_{cr}\approx0.9$) above which a singularity develops, which corresponds to a flow towards an arbitrary negative $X$. Moreover the evolution from odd data is much faster, which means that the zero mode is excited to a much higher acceleration.
The general rule is that even initial data have to be sufficiently localized to develop a singularity. Other initial data decay into a pair of defects. 

Thus, we may conclude that, starting from a perturbed $\phi=-1$ BPS solution, it is dynamically preferred to move in the positive direction, $X>0$, that is to form the kink-antikink pair, rather than pass to the negative direction, $X<0$, which results in a flow to an arbitrary negative field value at the origin. 

\subsection{Perturbation scheme}
To understand the results we have expanded the solution into powers of $A$ and considered the perturbation scheme
\begin{equation}
 \phi=-1+A\phi^{(1)}(x,t)+A^2\phi^{(2)}(x,t)+\cdots
\end{equation} 
In $\mathcal{O}(A)$ the equation is the well-known linear equation we studied earlier with the P\"oschl-Teller potential \cite{BPS-imp-solv}
\begin{equation}
 \phi^{(1)}_{tt}-\phi^{(1)}_{xx}+V(x)\phi^{(1)}=0.
\end{equation} 
The solution can be expressed  as a linear combination of the zero mode $\eta_0(x)=\sech^jx$ and some other modes $\eta_{rad}$ including radiation and (if any) higher bound modes.
\begin{equation}
 \phi^{(1)}=\alpha^{(1)} \eta_0+\eta^{(1)}_{rad}, \;\;\; \alpha^{(1)} =\frac{1}{N}\int_{-\infty}^\infty\sech^{j+\lambda}x\,dx, 
\end{equation} 
where $N$ is a normalization factor
\begin{equation}
 N^2=\int_{-\infty}^\infty\sech^{2j}x\,dx .
\end{equation} 
All the above integrals can be expressed in terms of hypergeometric functions. Note that in the case of odd initial data, the excitation of the zero mode is zero because the initial conditions are orthogonal to the zero mode.

The second order perturbation theory takes the form
\begin{equation}
 \phi^{(2)}_{tt}-\phi^{(2)}_{xx}+V(x)\phi^{(2)}= j\left(3j-\frac{1+3j}{\cosh^2x}\right)  {\phi^{(1)}}^2.
\end{equation}
Assuming that the second order solution can be decomposed as 
\begin{equation}
 \phi^{(2)}=\alpha^{(2)}(t)\eta_0+\eta^{(2)}_{rad}
\end{equation} 
we can project the above equation onto the zero mode $\eta_0$ and obtain the acceleration of the amplitude coming from the second order

\begin{equation}
 \frac{d^2\alpha^{(2)}}{dt^2}=\frac{j}{N}\int_{-\infty}^{\infty}\!\!dx\,\left(3j - \frac{1+3j}{\cosh^2x}\right) \frac{{\phi^{(1)}}^2}{\cosh^jx}.
\end{equation} 
The sign of the acceleration is governed by the sign of the bracket. In fact, it is mostly positive except for a small region around $x=0$. For $t=0$, the first order solution can be taken as our initial data $\phi^{(1)}=1/\cosh^\lambda x$ in the even case and 
$\phi^{(1)}=\sinh x / \cosh^\lambda x$ in the odd case. For larger values of  $\lambda$, the initial data are better localized and the negative part gives a larger contribution. In the odd initial data case, $\lambda$ would have to be much larger because $\phi^{(1)}(x=0)=0$. For wider initial conditions (smaller $\lambda$), the acceleration is positive.  Of course, in order to have a full description we would have to solve the full first order equation (including the radiation part) and then put it into the integral. We could partially assume that the radiation part would be radiated out quickly enough so that only the contribution from the zero mode remains. This simplification is in qualitative agreement with the numerical results described in the above subsection. This can also be justified by the observation that the radiation is usually less localized than the zero mode and becomes even less localized in time, giving only positive contributions. 

Concluding, most perturbations result in the evolution towards the creation of the kink-antikink pair rather than towards the singularity $(X\rightarrow -\infty)$. This is the key ingredient in understanding the existence of the vacuum wall. Namely, as the initially perturbed kink-antikink solution evolves, we approach the constant BPS solution with the original perturbation on top of it. As we know, a generic perturbation of the $\phi=-1$ solution leads to the creation of a kink-antikink pair which tends to increase their mutual distance. As a consequence, we observe that the initial solitons are reflected by the vacuum wall at $\phi=-1$. 

\section{Higher order spectral walls and bouncing solutions}
The spectral walls form when a discrete mode with frequency $\omega$ enters the continuous spectrum. However, each discrete mode may give rise to the appearance of its higher harmonics. Then, one can ask the question what happens if such higher harmonics hit the mass threshold 
\be
n \omega_n = E_{continuum} \;\;\; \Rightarrow \;\;\; \omega_n = \frac{j^2}{n} .
\ee
Of course, this occurs before the spectral wall. An obvious requirement for the existence of this effect is to excite a mode which is deep in the discrete spectrum, i.e., located in the vicinity of $\omega=0$. In our self-dual impurity $\phi^4$ model there is such a mode, namely the asymmetric superposition of the translation modes of the infinitely separated initial solitons. In other words, our initial configuration should be a kink-antikink BPS pair slightly shifted with respect to the origin and then boosted towards each other with a low velocity. Due to the fact that this mode has initially almost 0 frequency, there is potentially a large number of harmonics which may go through the mass threshold. However, higher harmonics have significantly suppressed amplitudes and in reality only a few of them are visible. 

During the evolution the higher harmonics enter the continuous spectrum. This happens for a certain value of the moduli space coordinate which is uniquely related to the value of the field at the origin. Again this defines a certain point in physical space (which for sufficiently large $X$ can be associated with the positions of the constituent solitons) which we will call {\it higher order spectral wall}. 

\begin{figure}
  \includegraphics[height=8.5cm]{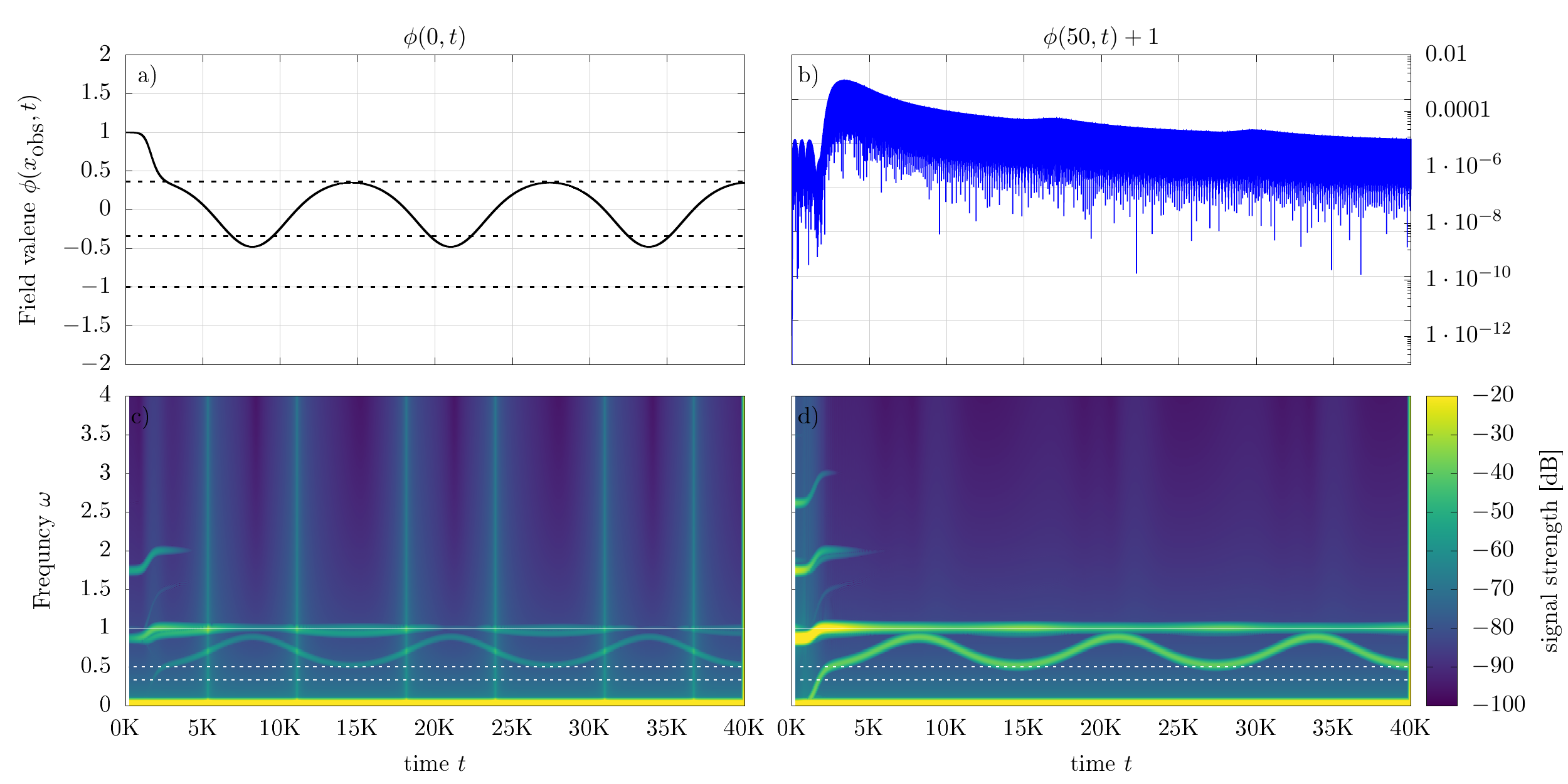}
\caption{Bouncing solution at the higher spectral walls for the impurity $\sigma_{j=1}$. Incoming solution with the asymmetric superposition of the shape modes excited. Initial velocity $v=0.005$. $A=0.0104$ and reflection at the second order spectral wall. }
\label{bouncing}
    \includegraphics[height=8.5cm]{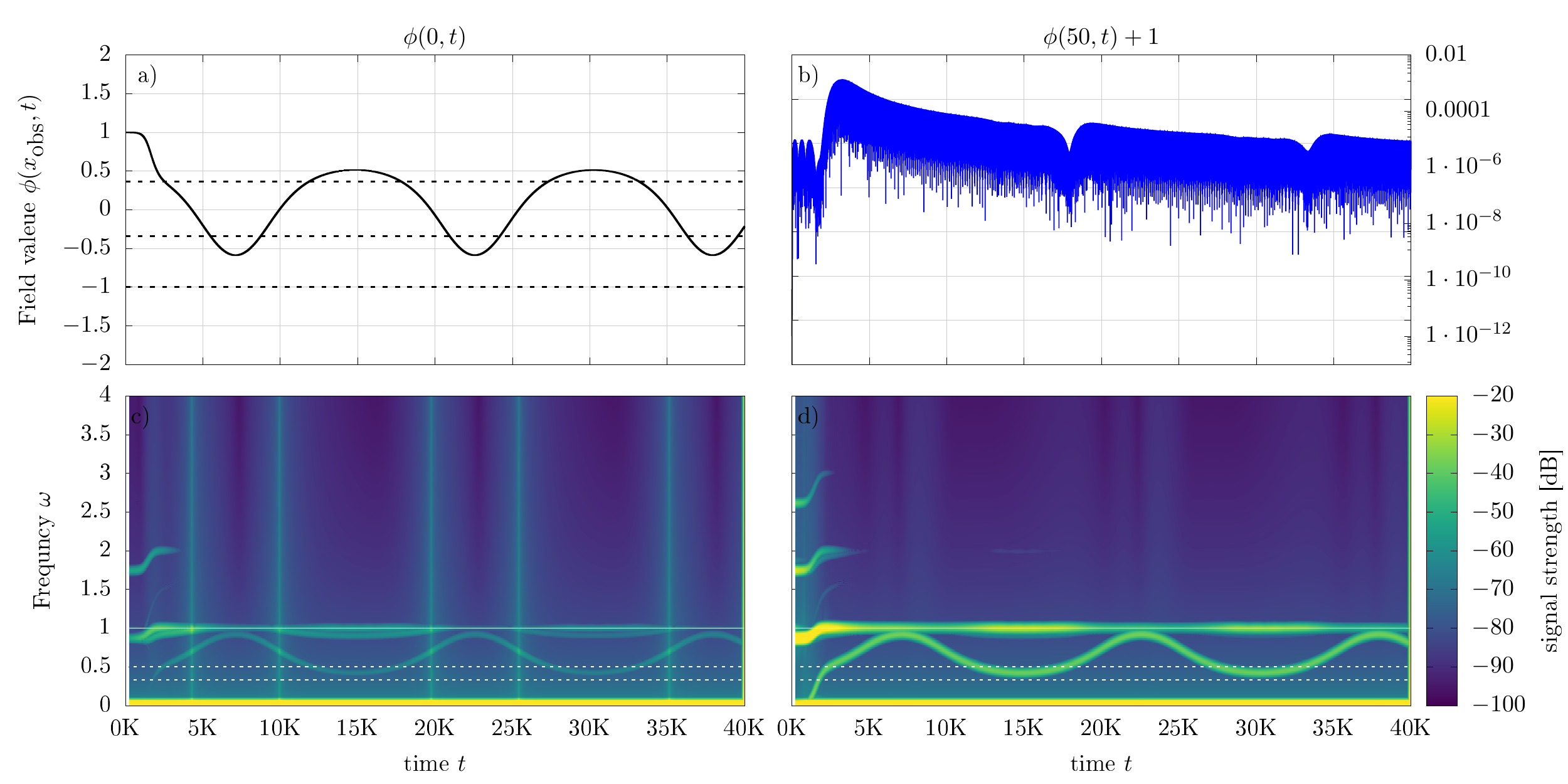}
\caption{Bouncing solution at the higher spectral walls for the impurity $\sigma_{j=1}$. Incoming solution with the asymmetric superposition of the shape modes excited. Initial velocity $v=0.005$. $A=0.01035$ and reflection at the third order spectral wall.}
\label{bouncing2}
\end{figure}

It is a matter of fact that the higher spectral wall phenomenon is the third condition which (beside the original spectral wall and the vacuum wall) stays behind the appearance of bouncing solutions. In order to clearly see this we present the structure of the exited modes during the collision of an incoming kink-antikink pair for which originally the asymmetric superposition of the shape modes was excited - see Fig. \ref{bouncing}. Such solutions are represented by some of the orange curves from Fig. \ref{asym-wall-plot} in Sec. III. In panel a) we show the field value at the origin as a function of time together with the spectral walls for all three discrete modes (horizontal dashed lines). In panel b) the field at $x=50$ is plotted which measures the amount of radiation emitted in the scattering process. Panels c) and d) show the signal strength, that is, the strength of exited frequencies at the origin and at $x=50$, respectively. This allows us to represent the strength of the symmetric as well as asymmetric modes during the time evolution of an excited  kink-antikink system. Here we plot the second and third order spectral walls.

Let us first explain the emergence of the bouncing solution in Fig. \ref{bouncing}. It shows a case with the mode amplitude $A=0.0104$ and the initial velocity of the solitons $v=0.005$. At the beginning only the asymmetric mode (asymmetric superposition of the shape modes) is excited. However, once the solution approaches the corresponding spectral wall at $\phi_0\approx 0.36$, the energy stored in this mode is transferred to another asymmetric mode, which here is the asymmetric  superposition of the translational modes. This happens because the system prefers to avoid the transition beyond the mass threshold. This asymmetric mode is originally very deep in the discrete spectrum with frequency only slightly above 0. While the solution is in the vicinity of the spectral wall, the frequency of the new asymmetric mode rises and approaches the second order spectral wall for this dynamically exited mode at $\phi_0=0.385$ - see Fig. \ref{bouncing} panel d) for $t \approx 3 000$. Observe that both spectral walls are located very close to each other. After a relatively long time at which the solution is bound by the walls it moves further and finally is reflected back by the vacuum wall (with probably some contribution of the spectral wall for this mode located in the same point $\phi_0=-1$). Then we observe a repetition of bouncing between the second order spectral wall (corresponding to this new asymmetric mode) and the vacuum wall. 

If the amplitude of the originally exited asymmetric mode (asymmetric superposition of the shape modes) is slightly smaller then we observe a reflection at the third order spectral wall of the dynamically exited asymmetric mode (asymmetric superposition of the translational modes) - see Fig.  \ref{bouncing2}. Hence now the trajectory moves slightly further toward $\phi_0=1$ which corresponds to the creation of a slightly more separated bouncing kink-antikink pair. 

\begin{figure}
    \includegraphics[height=8.0cm]{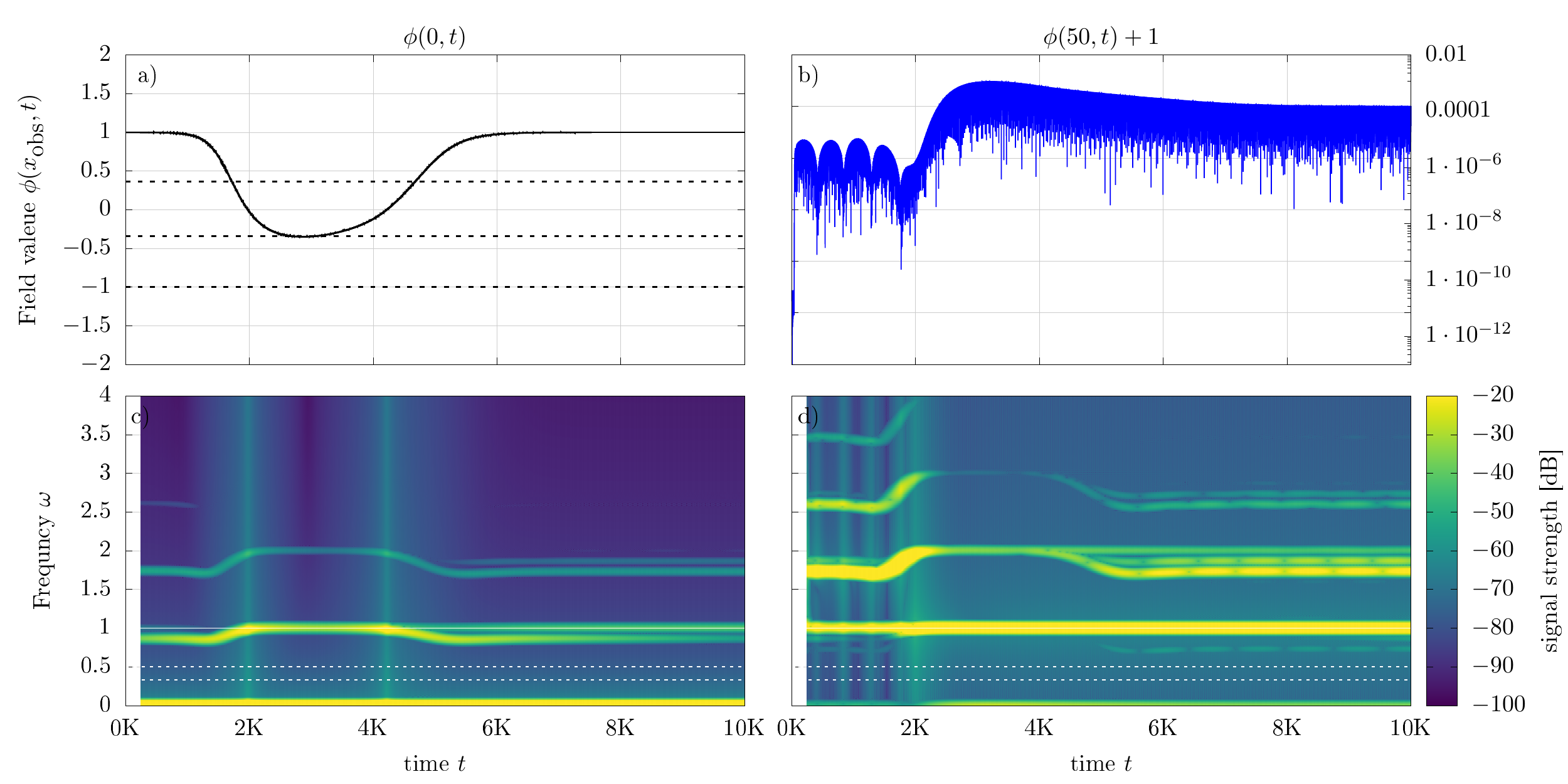}
\caption{Repulsion from the symmetric mode spectral wall  impurity $\sigma_{j=1}$. Incoming solution with the symmetric superposition of the shape modes excited. Initial velocity $v=0.005$. $A=0.0104$. }
\label{sym-wall}

\vspace*{0.5cm}

      \includegraphics[height=8.0cm]{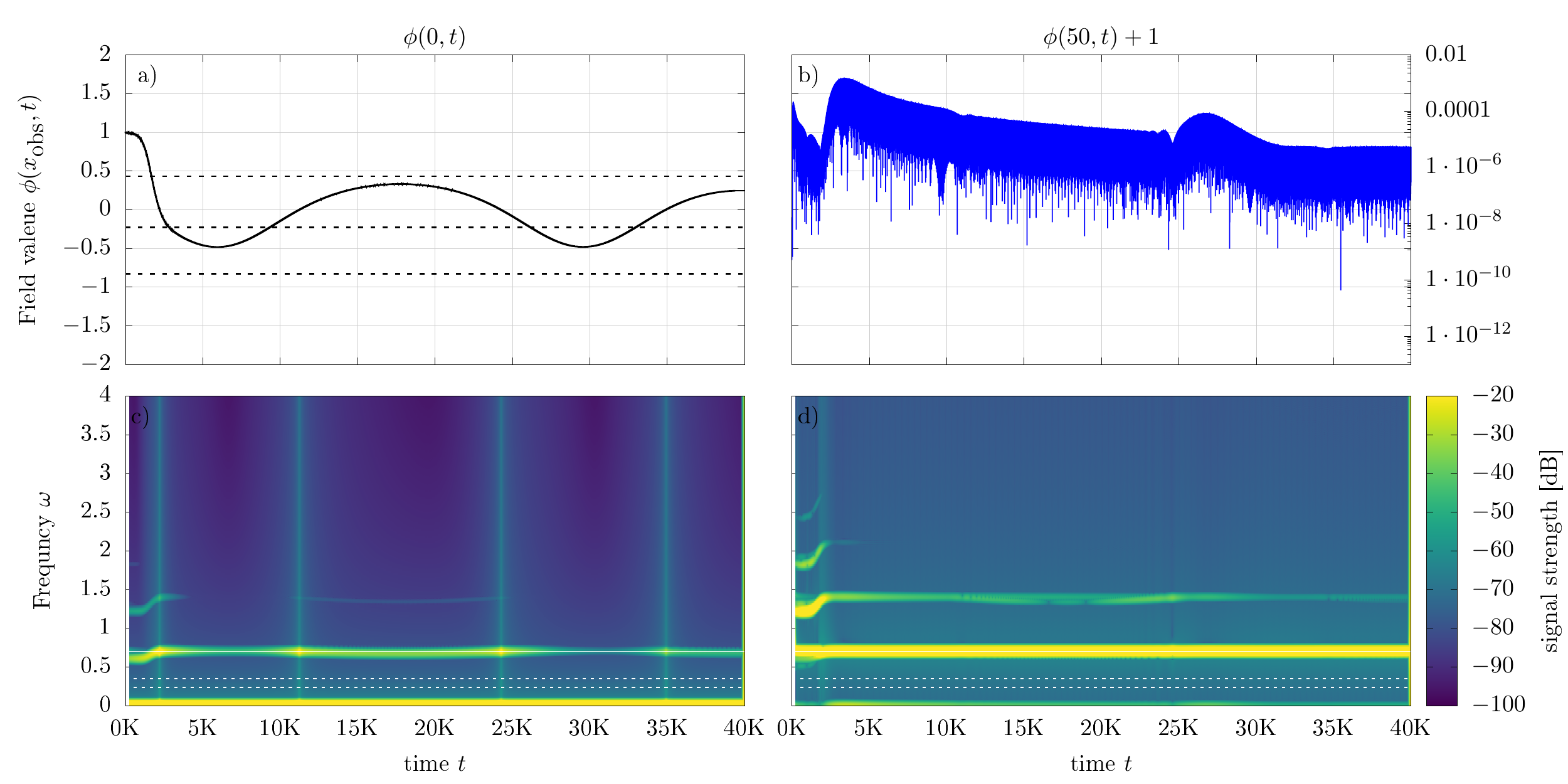}
\caption{Bouncing solutions impurity $\sigma_{j=0.7}$. Incoming solution with the symmetric superposition of the shape modes excited. Initial velocity $v=0.005$. $A=0.01012$.}
\label{sym-bouncing}
\end{figure}

However, there may exist another mechanism which also contributes to the formation of the bouncing structures. Let us consider a BPS solution with the symmetric superposition of the shape modes excited. As we know, the relevant spectral wall in this case is less visible due to the impact from the vacuum wall. Nonetheless, such a wall is easy to recognise in Fig. \ref{sym-wall}. For a bouncing solution ($j=0.7$ and $A=0.01012$) we do not find any higher spectral wall responsible for this behaviour, see Fig. \ref{sym-bouncing}. It looks rather as if an effective potential emerges which prevents the system from forming an infinitely separated pair of kink and antikink. This problem requires further studies. 

\vspace*{0.2cm}

We conclude that, even in our simplified model where the static intersoliton force vanishes, the emergence of the bouncing solutions seems to possess a rather involved mechanism which requires the existence not only of the spectral walls but also the vacuum wall and higher order spectral walls. Furthermore, an additional, not yet explained mechanism can also play a role. 

\vspace*{0.2cm}

By a closer inspection of Fig.  \ref{bouncing} and Fig.  \ref{bouncing2} we can see another qualitative difference between these two scattering scenarios. Namely, there are periodic suppressions and enhancements  of the radiation - see panel b) in Fig. \ref{bouncing2}. This phenomenon is absent in Fig.  \ref{bouncing}. The source of this effect is the transition of the solution through the higher spectral walls, here through the second order spectral wall. We analyze this phenomenon in the next section. 
\section{Higher order spectral walls and radiation bursts}

\begin{figure}
  \includegraphics[height=8.5cm]{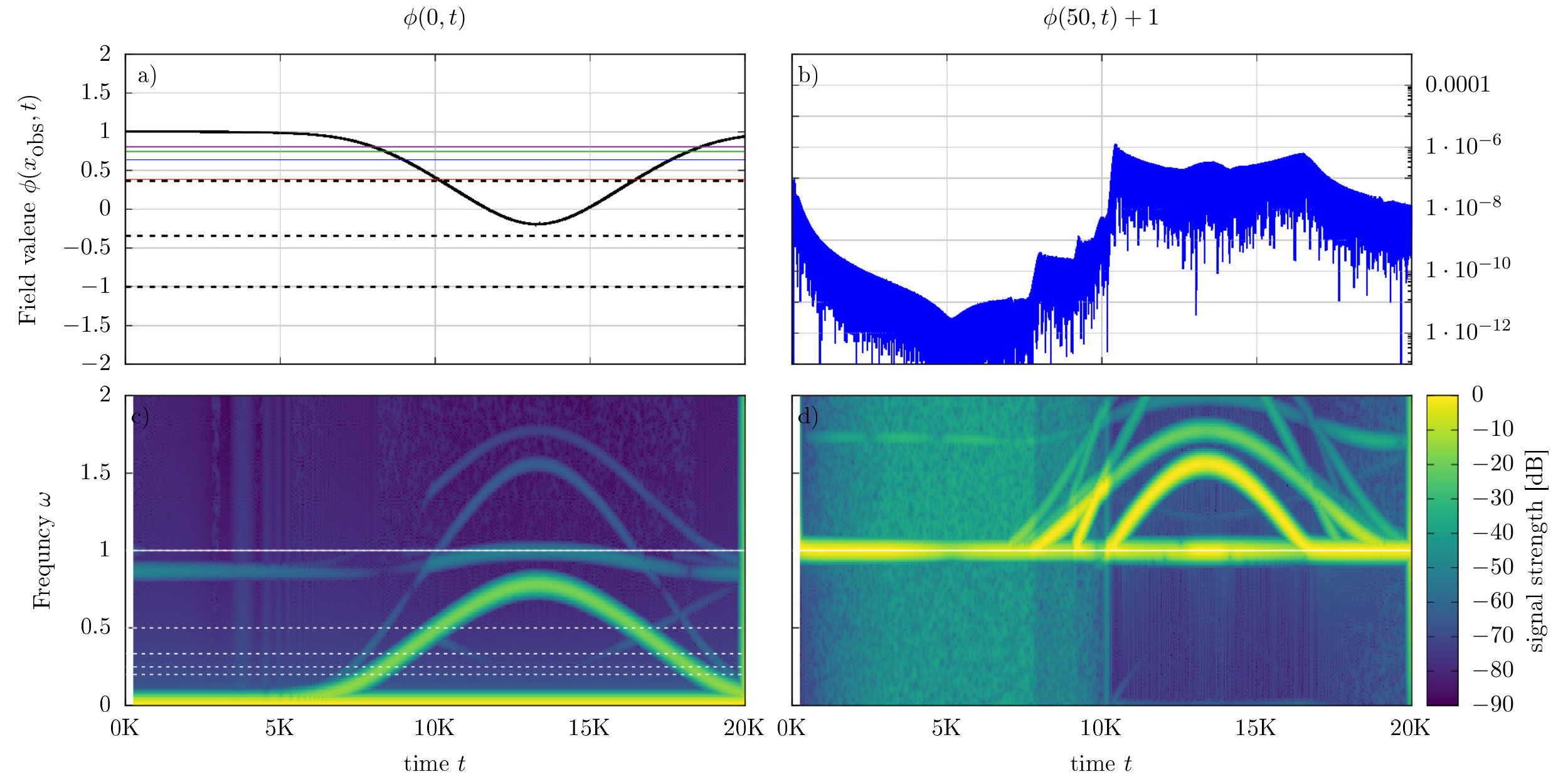}
\caption{Radiation at the higher spectral walls for impurity $\sigma_{j=1}$. Initial distance $2x_0=20$, displacement $0.05$ and velocity $v=0.001$.}
\label{radiation-plot1}
\end{figure}

\begin{figure}
  \includegraphics[height=8.5cm]{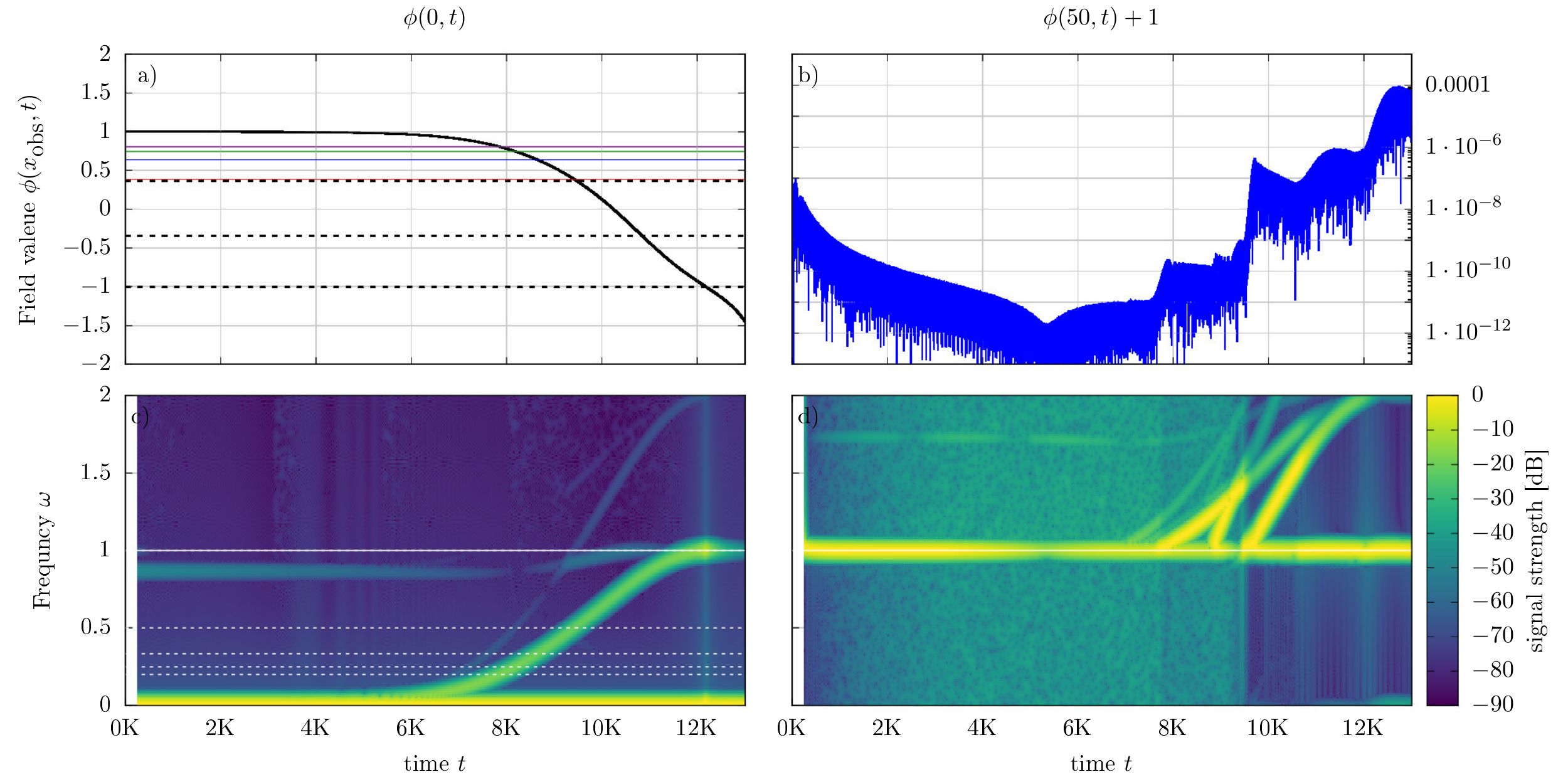}
\caption{Radiation at the higher spectral walls for the impurity $\sigma_{j=1}$. Initial distance $2x_0=20$, displacement $0.03$ and velocity $v=0.001$.}
\label{radiation-plot2}
\end{figure}

In Fig. \ref{radiation-plot1} we present a collision of the kink and antikink with the asymmetric superposition of the translation modes excited for the impurity with $j=1$. As we previously commented this mode is very deep in the discrete spectrum which allows for the existence of many higher order spectral walls. Specifically, the initial state is given by a kink-antikink pair at a distance $2x_0=20$ and with a displacement with respect to the origin of $0.05$. The initial velocity of the solitons is $v=0.001$. In panel a) we show the field value at the origin as a function of time together with the first few higher order spectral walls for this mode (horizontal color lines). Also the spectral walls for all three discrete modes are marked (horizontal dashed lines). In panel b) the field at $x=50$ is plotted. Panels c) and d) show the signal strength, that is, the strength of exited frequencies at the origin and at $x=50$, respectively. To avoid any dependence of the results for the field strength on some accidental symmetry, we use the quantity $\sqrt{\phi^2+\phi_x^2}$ for the Fourier analysis. 

In this case, the excited mode is so strong that the solution is repelled before the mode enters the continuous spectrum. Hence, the constituent solitons scatter back before the corresponding spectral wall is met. However, it is clear that the second harmonic crosses the mass threshold. Interestingly, just before this happens, the second harmonic interacts with another mode, that is, the symmetric superposition of the shape modes. This mode is initially exited due to nonlinearities of the model. As a consequence, this symmetric mode is enhanced, which can contribute to some additional repulsion of the solution close to its spectral wall at $\phi_0=-0.343121$. So, higher harmonics may transfer energy between the modes. Next, after the solution scatters back, the second harmonic enters the discrete spectrum. Thus, higher (here the second order) spectral walls can appear in pairs.  Note that from the point of view of the signal strength, higher harmonics are much better visible in a far distance region, here at $x=50$, 
after crossing the mass threshold, see Fig. \ref{radiation-plot1}, panel d), where third and fourth harmonics show up. 

\begin{figure}
  \includegraphics[height=8.5cm]{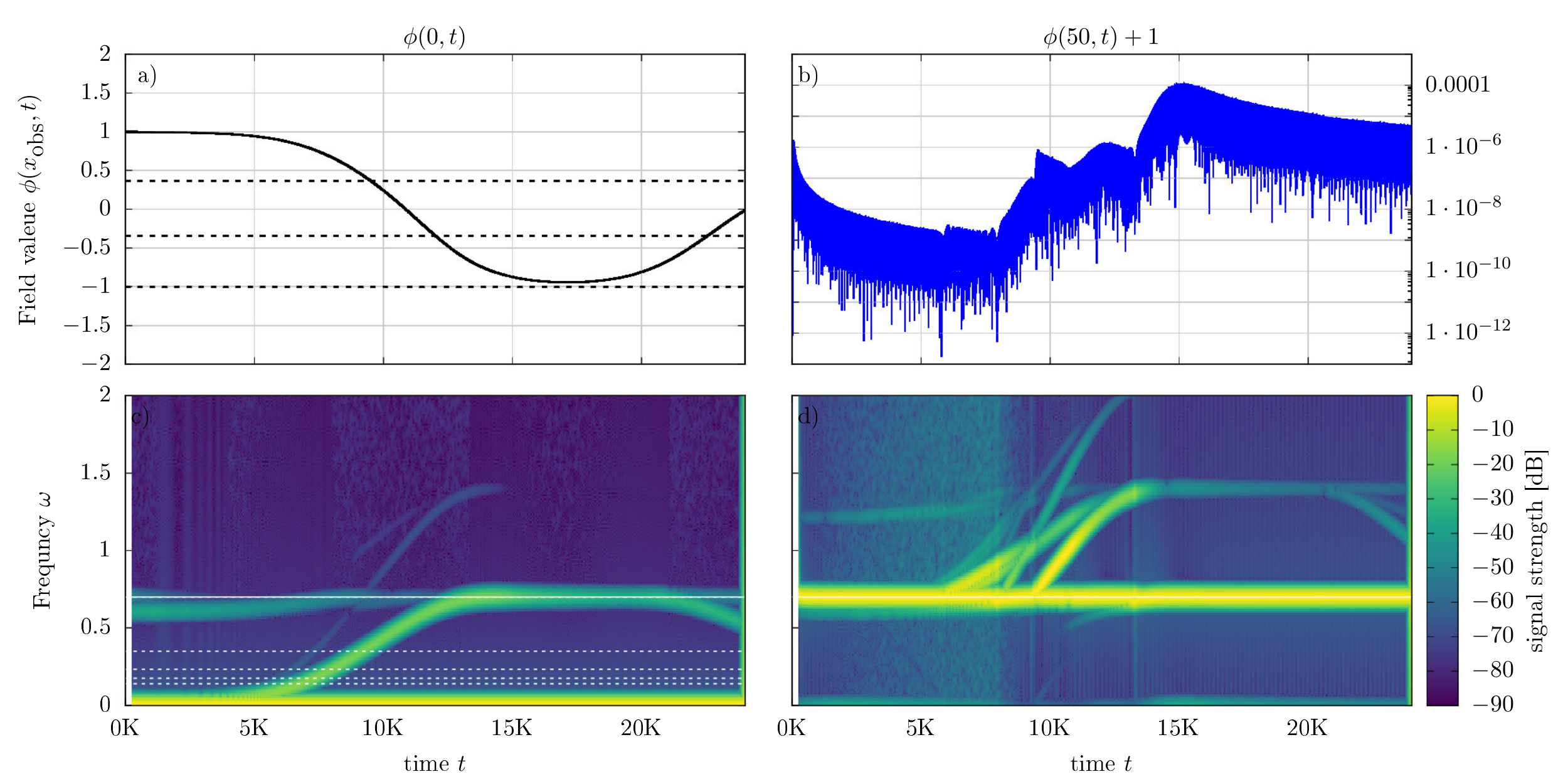}
\caption{Radiation at the higher spectral walls for the impurity $\sigma_{j=0.7}$. Initial distance $2x_0=20$, displacement $0.028$ and velocity $v=0.001$.}
\label{radiation-plot3}
\end{figure}

As a matter of fact, the most spectacular evidence for the higher spectral walls is provided by radiation. Each transition through higher spectral walls leads to a huge emission of radiation. In fact it is a very violent process as we clearly see even the fifth order higher spectral wall. The biggest enhancement of radiation occurs during the crossing of the second order spectral wall. Note that such  radiation bursts can detect a spectral wall much before this wall is actually met. 

In Fig. \ref{radiation-plot2} we analyze the same situation but with a smaller displacement $0.03$. This results in a smaller amplitude of the excited deep mode  which finally allows the configuration to go through the corresponding spectral wall as well as through the vacuum wall (located at the same point). Again we see several radiation bursts due to the higher order spectral wall. Furthermore, the transition through the vacuum wall also gives rise to another huge radiation burst. At this point, the solution reduces the amplitude of the excited mode significantly and can pass over the vacuum wall.  Note an enormously large enhancement of radiation from $10^{-12}$ to $10^{-4}$, that is, {\it eight} orders of magnitude.  
\begin{figure}
  \includegraphics[height=8.5cm]{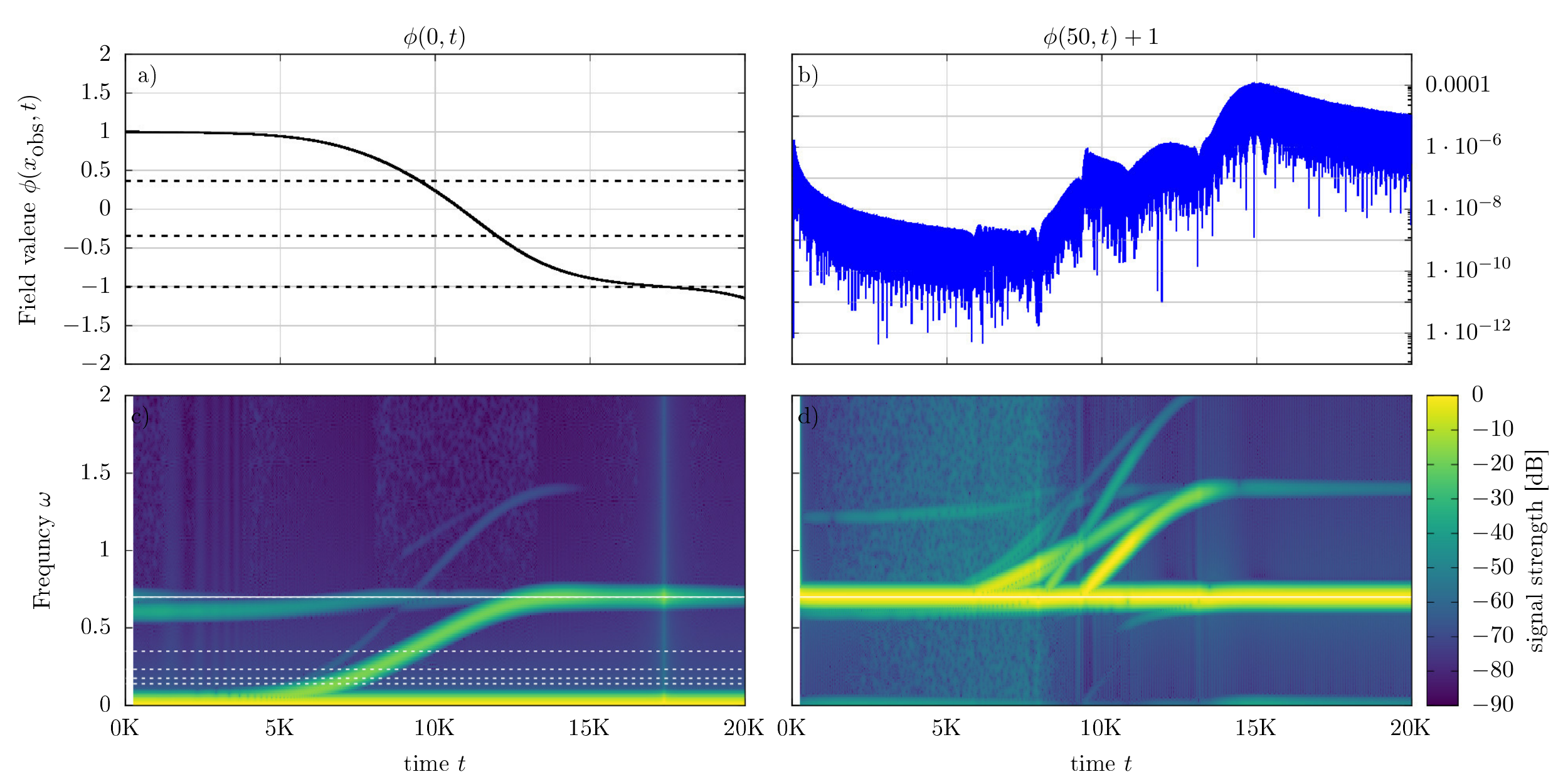}
\caption{Radiation at the higher spectral walls for impurity $\sigma_{j=0.7}$. Initial distance $2x_0=20$, displacement $0.0279$ and velocity $v=0.001$.}
\label{radiation-plot4}
\end{figure}

In Fig. \ref{radiation-plot3}  we plot the case with the impurity with $j=0.7$. Here the relevant spectral wall does not coincide with the vacuum wall and is located at $\phi_0= -0.827630$. The initial displacement is $0.028$, $x_0=10$ and the initial velocity is $v=0.001$. The amplitude of the  mode is such that the solution goes through the spectral wall and scatters back at the vacuum wall. Strictly speaking, the solution is trapped between these walls for quite a long time during which it radiates by a quasinormal mode. Obviously, this resonant mode originates from the originally excited mode after its transition through the mass threshold at the spectral wall. The resonance is rather narrow as we are all the time close to the pertinent spectral wall. 

Finally, in Fig.  \ref{radiation-plot4}  we again demonstrate the case with $j=0.7$, but now the displacement is $ 0.027900$. The amplitude of the mode is smaller and the solution passes through all walls, the vacuum wall included. The radiation bursts at the (higher) spectral walls are again very well visible. Once the solution passes the vacuum wall, the perturbation (localized on the deepest bound mode) is reduced, which allows the system to move towards the $X<0$ side of the moduli space. 

\section{Summary}
In this paper, we have applied the framework of the self-dual soliton impurity models to study kink-antikink scattering in $\phi^4$ theory in the limit when there is no static force between the scattering solitons. This means that a kink-antikink pair is a BPS (self-dual) solution of a pertinent Bogomolny-type equation and, therefore, saturates the corresponding topological energy bound. Furthermore, there is a whole family of such solutions which span a one-dimensional moduli space with a coordinate which, to some extent and with certain limitations \cite{BPS-imp-solv}, may be identified with the distance between the scattering solitons. Then the lowest order approximation to the annihilation process, which reproduces the full dynamics if the solitons have small velocity and no internal modes are excited, is provided by the gradient flow on the moduli space. 

However, the main aim of our work was to study the regime beyond the geodesic approximation, i.e., systematically analyze the impact of excitations of the internal modes on the scattering process. As there is no static inter-soliton force, the solitons only interact via excited modes and radiation. Thus we have the unique possibility to disentangle the role of the modes in the annihilation process. We emphasize again that we perform the computations in a mathematically rigorous manner, as the BPS static solutions exist for any kink-antikink separation distance. Thus, the spectral structure is well defined at any stage of the collision.   

As we expected, the excitation of bound modes (minimally 3 modes) gives rise to spectral walls which are well observed in the full numerics. The spectral walls are defined by the position on the moduli space at which the corresponding bound mode enters the continuum spectrum. It results in a space-localized region at which excited solitons interact nontrivially, that is, can be reflected, feel a temporary attraction or go through. The actual interaction depends on the amplitude of the excited mode. 

Although the spectral wall is a selective phenomenon, which means that each mode feels its own spectral wall, sometimes a significant excitation of other modes can be observed while we move through the wall. This occurs for example in the case when the kink-antikink solution with the initially asymmetric mode excited (asymmetric superposition of the shape modes) hits its spectral wall which results in the dynamical excitation of another asymmetric mode (asymmetric superposition of the translational modes), as discussed in Sec. VI. In a sense, the system always prefers to avoid the situation when a mode passes the mass threshold. If it is possible to transfer the energy to other modes which are not related to the actual spectral wall, then this is what happens. 

\hspace*{0.2cm}

A completely novel finding is the discovery of the {\it vacuum wall}. This wall is defined by a quite ordinary point on the moduli space, which corresponds to the $\phi=-1$ lump solution, and it does not cause any special effects in the geodesic approximation or the spectral structure flow. However, exactly at this point the system has zero static energy density and, therefore, it is very sensitive to any initially added perturbation. We found that a generic perturbation prefers to create a kink-antikink pair rather than flow to the singularity (of the moduli space metric). Effectively, this leads to the appearance of a repulsive interaction if a perturbed BPS solution approaches the point $X=0$ on the module space (i.e., $\phi_0=-1$). 
Hence, this wall acts in a way similar to a spectral wall, with the important difference that it is not related to any particular internal mode. It repels, temporary confines or transmits a kink-antikink solution with {\it any} mode excited. When a spectral wall is located too close (on the moduli space) to the vacuum wall, then its impact on the solution is less visible. As a consequence, such a spectral wall can be less visible or even completely hidden in the shadow of the vacuum vacuum wall. Undoubtedly, the vacuum wall strongly influences the behavior of solutions close to this spectral wall and a geodesic trajectory ceases to be an efficient observable for the detection of such a spectral wall. 

It is important to stress that both types of walls have, to some extent, a qualitatively similar impact on the incoming solutions. However, they originate from completely different mechanisms. Furthermore, owing to its nonselective nature, we expect that the vacuum wall affects soliton dynamics in a more profound way. This observation was possible exclusively due to the BPS nature of our model, which led to a sort of solvability. 

\hspace*{0.2cm}

The second novel phenomenon is the existence of the {\it higher order spectral walls}. They are higher harmonics counterparts of the usual spectral walls, which occur when a higher harmonic (of a bound mode) enters the continuous spectrum. Again, at the locus of the higher order spectral wall a kink-antikink solution with a pertinent mode excited can be repelled or pass through, although its impact on the solution seems to be in general weaker that the original spectral wall.   

The higher order spectral walls are, on the other hand, very clearly visible in enhanced radiation ({\it radiation bursts}). Notably, the cumulative effect of passing through several higher order spectral walls may be an increase in the amount of radiation by eight order of magnitudes, which might make them interesting phenomena from a phenomenological point if view and even lead to some applications.  Equivalently, we can say that the radiation bursts are very sensitive observables allowing to detect spectral walls before they are met. By means of the radiation bursts, we can easily see spectral walls even if they are hidden in the shadow of the vacuum wall and not visible in a direct soliton observation. 

\hspace*{0.2cm}

The last important result is the existence of {\it bouncing solutions}. Such a configuration represents an oscillating (bouncing) kink-antikink pair formed during the scattering of an asymptotically free kink and antikink with only one mode excited. We found that this class of solutions is trigged by three ingredients: the spectral wall (related to the excited mode), the vacuum wall (which turns back the trajectory to a kink-antikink pair) and the higher order spectral walls (related to another dynamical excited mode). Moreover, we observed bouncing trajectories where the reflection at the higher spectral walls should be replaced by another, not yet clearly identified, mechanism. It is quite surprising that even in such a simplified theory, with no static force between solitons, the bouncing states are formed by a rather involved mechanism. We underlined that our  (partially quantitative, partially qualitative) explanation was possible only due to the BPS nature of the model. 

\hspace*{0.2cm}

A more general comment is that the distortion of a bound mode, and especially its transition to the continuous spectrum, can have a huge impact on the soliton dynamics. This is a very important observation, as this fact is not captured by the collective coordinates approach where bound modes are assumed to exist on the whole unstable manifold. Hence, this can be the crucial source of the failure of the effective models, which undoubtedly have to be corrected. 

A further general remark is that if a system possesses several spectral walls (bound modes) then the resulting kink-antikink dynamics may be very complicated. 

\vspace*{0.2cm}

As we mentioned at the beginning, this work is a first step in the program of a systematic understanding of the role of the modes, static forces and radiation in the kink-antikink scattering in the $\phi^4$ model. Therefore, an obvious continuation of our research requires the inclusion of a self-duality (BPS) breaking term, which effectively introduces a static inter-soliton force. Strictly speaking, this would destroy the moduli space. However, as it can be performed in a perturbative manner, we can still be in a near-BPS regime. Hence, the (near) geodesic approximation should still be meaningful, provided by an unstable moduli manifold \cite{manton-unstabl}. As a consequence, it could be possible to verify how the static force influences the phenomena found in the current paper. Do they survive in a non-BPS system? Do they give rise to some other phenomena? Eventually, this could give a chance to explain the origin of the dynamical properties of the pure $\phi^4$ theory. Here, the recent results on the construction of the improved moduli space of kink-antikink scattering in $\phi^4$ theory may be very helpful \cite{nick-we}.

Next, the presented framework can be applied to the analysis of scattering of kinks in various models in (1+1) dimensions. Let us mention deformed $\phi^4$ theory \cite{phi4-deformed}-\cite{zhong}, hyperbolic models \cite{bazeia}, $\phi^6$ theory \cite{gani-phi6}-\cite{kev-phi6}, so-called long tail soliton models \cite{long}-\cite{nick-long-tail} as well as scalar theories with two \cite{izquierdo-1}-\cite{izquierdo-3} or even more \cite{gani-three}-\cite{wojtek} fields.

Of course, in a parallel line of investigation, one should search for the effects presented here (spectral walls, higher order spectral walls and radiation bursts) in other self-dual solitonic systems. Natural candidates seem to be self-dual models in higher dimensions, where the BPS sector contains multi-soliton configurations which posses a nontrivial spectral structure flow on the moduli space. Let us mention the Abelian Higgs model at critical coupling or self-dual monopoles as two obvious examples.

\end{document}